\documentclass[
  a4paper, twocolumn, pra, superscriptaddress, amsmath, amssymb, mathrsfs,
  showpacs
]{revtex4-1}
\usepackage{graphicx}
\usepackage{epstopdf}
\usepackage[table]{xcolor}
\usepackage{braket}
\usepackage{algorithm}
\usepackage{algorithmicx}
\usepackage{algpseudocode}
\usepackage[colorlinks]{hyperref}
\usepackage{amsthm}
\usepackage{mathtools}
\usepackage{siunitx}
\usepackage{setspace}
\usepackage{tikz}
\usetikzlibrary{shapes, arrows, positioning, calc, fadings}
\usetikzlibrary{decorations.pathmorphing}
\usetikzlibrary{backgrounds,fit,decorations.pathreplacing}

\colorlet{alert}{red!80!black}
\definecolor{SchoolColor}{rgb}{0.9608, 0.5020, 0.1451} 
\definecolor{chaptergrey}{rgb}{0.9608, 0.5020, 0.1451}

\hypersetup{%
  linkcolor=blue,
  citecolor=blue,
  filecolor=black,
  menucolor=black,
  urlcolor =black
}

\newcommand{\im}{\ensuremath{\textup{i}}}
\newcommand{\ci}{\im}

\newcommand{\rp}{\ensuremath{\mathfrak{Re}}}
\newcommand{\ip}{\ensuremath{\mathfrak{Im}}}

\newcommand{\op}[1]{\ensuremath{\boldsymbol{\mathsf{\hat{#1}}}}}

\newcommand{\lop}[1]{\ensuremath{\boldsymbol{\mathsf{\mathcal{#1}}}}}

\newcommand{\pulse}{\mathcal{E}}

\newcommand{\dd}{\ensuremath{\mathrm{d}}}

\begin{document}

\title{%
  Dissipative Quantum Dynamics and Optimal Control using Iterative Time
  Ordering: An Application to Superconducting Qubits
}

\author{Daniel Basilewitsch}
\thanks{These two authors contributed equally}
\affiliation{Theoretische Physik, Universit\"{a}t Kassel,
Heinrich-Plett-Stra{\ss}e 40, D-34132 Kassel, Germany}

\author{Lutz Marder}
\thanks{These two authors contributed equally}
\affiliation{Theoretische Physik, Universit\"{a}t Kassel,
Heinrich-Plett-Stra{\ss}e 40, D-34132 Kassel, Germany}

\author{Christiane P. Koch}
\email{christiane.koch@uni-kassel.de}
\affiliation{Theoretische Physik, Universit\"{a}t Kassel,
Heinrich-Plett-Stra{\ss}e 40, D-34132 Kassel, Germany}

\date{\today}
\begin{abstract}
  We combine a quantum dynamical propagator that explicitly accounts for quantum
  mechanical time ordering with optimal control theory. After analyzing its
  performance with a simple model, we apply it to a superconducting circuit
  under so-called Pythagorean control. Breakdown of the rotating-wave
  approximation is the main source of the very strong time-dependence in this
  example. While the propagator that accounts for the time ordering in an
  iterative fashion proves its numerical efficiency for the dynamics of the
  superconducting circuit, its performance when combined with optimal control
  turns out to be rather sensitive to the strength of the time-dependence. We
  discuss the kind of quantum gate operations that the superconducting circuit
  can implement including their performance bounds in terms of fidelity and
  speed.
\end{abstract}
\maketitle

\section{Introduction}
\label{sec:intro}
The interaction of matter with electromagnetic fields provides access to study
the structure and dynamics of quantum systems. Quantum control takes this
concept one step further, asking how external fields can be used to steer
the dynamics in a prespecified, desired way. Optimal control
theory~\cite{GlaserEPJD15,WerschnikJPB07} is a set of methods to derive the
shape of the electromagnetic fields that accomplish a given task in the best
possible way. Its application ranges from nuclear magnetic
resonance~\cite{KhanejaJMR05}, driven electron
dynamics~\cite{CastroPRL12,HellgrenPRA13,GreenmanPRA15,GoetzPRA16}, or
photoinduced chemical reactions~\cite{SomloiCP93} all the way to quantum
information science~\cite{GoerzJPB11,WattsPRA15,GoerzPRA15} (see
Ref.~\cite{GlaserEPJD15} for a more comprehensive overview). Very often,
optimal control calculations yield pulse shapes which vary very strongly as
a function of time. This results in non-negligible effects of quantum mechanical
time ordering, due to the non-commutativity of an explicitly time-dependent
Hamiltonian with itself at two different instances of time.

While one might naively argue that these effects vanish for sufficiently small
time steps, such an argument overlooks the accumulation of error that
accompanies the partitioning of a given overall propagation time into ever
smaller steps. In contrast, a propagation method that \textit{explicitly}
accounts for time ordering will allow to accurately assess the impact of time
ordering on quantum optimal control. At the same time, the propagation scheme
needs to be numerically efficient since optimal control algorithms require many
propagations to derive suitable pulse shapes. Generally, semi-global methods
offer the best compromise between accuracy and efficiency. They are based on
splitting the overall domain of integration into small sub-intervals and solving
the differential equation of interest with a spectral method, i.e., a global
approximation, within the small interval. In the spatial domain, these
approaches are typically summoned under the term finite element discrete
variable representation. An equivalent approach in the time domain is given by
the iteratively time ordering propagator introduced in
Ref.~\cite{ndong2010chebychev}. In a nutshell, it is based on rewriting the
action of the explicitly time-dependent part of the Hamiltonian onto the system
state as an inhomogeneity such that the homogenous Schr\"odinger equation with
explictly time-dependent Hamiltonian becomes an inhomogeneous Schr\"odinger
equation with time-independent Hamiltonian. The formal solution of the resulting
inhomogeneous Schr\"odinger equation can be determined using a spectral
method~\cite{NdongJCP09}, similarly to the Chebyshev propagator
\cite{tal1984accurate} for the (homogeneous) Schr\"odinger equation with
time-independent Hamiltonian. Such a propagator that explicitly accounts for
time ordering is capable of efficiently computing arbitrary time-dependencies of
the quantum system up to a very high precision \cite{ndong2010chebychev}. This
is in contrast to more commonly used propagation methods such as the Chebyshev
propagator with a piecewise constant (PWC) approximation of the Hamiltonian's
time-dependence. Moreover, the iterative time ordering approach is not limited
to the conventional time-dependent Schr\"odinger equation (TDSE) but can also be
extended to other forms of the equation of motion, be it non-linear or
non-unitary \cite{tal2012new,schaefer2017semi}.

Having a highly accurate while still efficient propagation method for explicitly
time-dependent problems at hand, we can assess the role of time ordering in
quantum optimal control theory. To this end, we merge the iteratively time
ordering propagator~\cite{ndong2010chebychev,tal2012new,schaefer2017semi} with
a gradient-based quantum optimal control algorithm, namely Krotov's method
\cite{Konnov.AutomRemContr.60.1427,SomloiCP93,SklarzPRA02,Palao.PRA.68.062308,
Reich.JCP.136.104103}. Compared with other optimal control variants, Krotov's
method comes with the advantage that monotonic convergence is guaranteed for
a wide range of control problems~\cite{Reich.JCP.136.104103}, including
non-linear equations of motion, a non-linear interaction with the control or
`unusual' target functionals that are useful in the context of quantum
information~\cite{WattsPRA15,GoerzPRA15}. Note that a combination of iterative
time ordering and quantum optimal control holds the promise of reducing the
dimension of the control search space, when using a PWC representation of the
external field. A smaller control space dimension should speed up convergence of
the local iterative optimization. We first test the new algorithm for a simple
control problem, the quantum harmonic oscillator under frequency control, and
then apply it to quantized superconducting circuits. These are one of the most
promising physical platforms for quantum information processing. Unlike qubits
encoded in atoms or ions, the rotating wave approximation is not well justified
in this case. In other words, the external fields driving the dynamics of
superconducting circuits often exhibit a very strong time-dependence. This makes
them a suitable test case for our algorithm.

The paper is organized as follows. We present the iteratively time ordering
propagator and its combination with Krotov's method for optimal control in
Sec.~\ref{sec:methods}. Section~\ref{sec:bench} analyzes the performance of the
propagator as well as its combination with Krotov's method for the harmonic
oscillator. The application of the new method to a superconducting circuit is
presented in Sec.~\ref{sec:pc}. We conclude in Sec.~\ref{sec:conclusions}.

\section{Numerical Methods}
\label{sec:methods}
We briefly recall the iterative time ordering (ITO) quantum
propagator~\cite{ndong2010chebychev,tal2012new,schaefer2017semi} in
subsection~\ref{subsec:methods:itoprop} and detail its implementation in
Appendix~\ref{app:implementation}. In subsection~\ref{subsec:methods:itoqoct},
we discuss how to combine iterative time ordering with Krotov's method for
quantum optimal control
\cite{Konnov.AutomRemContr.60.1427,Reich.JCP.136.104103}.

\subsection{Quantum Dynamics with Iterative Time Ordering}
\label{subsec:methods:itoprop}
The basic idea of the ITO quantum propagator is to rewrite the equation of
motion as an inhomogeneous first order ordinary differential equation (ODE). The
formal solution of this ODE can be constructed by expansion into orthogonal
polynomials, similar to the Chebychev \cite{tal1984accurate} or Newton
propagator \cite{RonnieReview94} for the standard time-dependent Schr\"odinger
equation with time-independent Hamiltonian. Since the inhomogeneity depends on
the solution of the ODE that one seeks, it must be determined iteratively in
a self-consistent loop.

No matter what is the specific physical system at hand, its time evolution is
described by a first order (in time) differential equation. Most commonly, the
equation of motion is linear in the state of the system. This is true for both
closed and open quantum systems. In the first case, the time-dependent
Schr\"odinger equation has to be solved,
\begin{eqnarray}\label{eq:tdse}
 \frac{\partial}{\partial t} \ket{\psi(t)}
  &=&
  - \frac{\ci}{\hbar} \op{H}(t) \ket{\psi(t)}
  =
  - \frac{\ci}{\hbar}\op{H}_0 \ket{\psi(t)} + \ket{\phi(t)},
\end{eqnarray}
where, for convenience, we have separated the explicitly time-dependent part,
\[
  \ket{\phi(t)} = - \frac{\ci}{\hbar}\op{W}(t) \ket{\psi(t)}.
\]
Similarly for open quantum systems, the Liouville von Neumann equation reads
\begin{eqnarray}\label{eq:LvN}
  \frac{\partial}{\partial t} \op{\rho} (t)
  &=& - \frac{\ci}{\hbar} \big[ \op{H}(t), \op{\rho}(t) \big] + \\
  && \sum_{k=1}^{N^2-1} \gamma_k \Big( \op{L}_k \op{\rho}(t)
   \op{L}_k^\dagger - \frac{1}{2} \big\{ \op{L}_k^\dagger \op{L}_k,
   \op{\rho}(t) \big\} \Big) \nonumber\\
   &=& \lop{\mathcal{L}}(t) \op{\rho}(t) \nonumber
  = \lop{\mathcal{L}}_0 \op{\rho}(t) + \op{\sigma}(t),
\end{eqnarray}
where the explicitly time-dependent part of the generator is captured by
$\op{\sigma}(t)$,
\[
\op{\sigma}(t) = \lop{V}(t) \op{\rho}(t),
\]
and where we have assumed the Lindblad form~\cite{Breuer.book}. However, the
equation of motion may also depend non-linearly on the state of the system.
Famously, it does so in time-dependent functional theory
\cite{RungePRL84,MarquesARPC04,TDDFT2012}. Another example is dynamics in the
mean-field approximation, such as the Gross--Pitaevskii equation in case of
a Bose-Einstein condensate,
\begin{eqnarray}
  \label{eq:GP}
  \ci \hbar \frac{\partial}{\partial t} \ket{\psi(t)}
  &=& \Big( \frac{\op{p}^2}{2 m} + \op{V} + g | \psi (t) |^2 \Big) \ket{\psi(t)}
  \\
  &=& \op{H}_0 \ket{\psi(t)} + \ket{\phi(t)}, \nonumber
\end{eqnarray}
Here, we have separated out the non-linear term, in analogy to the explicit
time-dependence in Eqs.~\eqref{eq:tdse} and \eqref{eq:LvN}
\[
  \ket{\phi(t)} = \op{W}(\psi,t) \ket{\psi(t)}.
\]
Equations~\eqref{eq:tdse} to \eqref{eq:GP} have in common that they can all be
written in the form of an inhomogeneous first order differential
equation~\cite{tal2012new},
\begin{align}\label{eq:methods:inhom_eom}
  \frac{\dd}{\dd t} u(t)
  =
  G(t) u(t)
  =
  G_0 u(t) + s \big( u(t), t \big).
\end{align}
Here, the operator $G_0$ acting on the state $u(t)$ is time-independent and the
inhomogeneity $s(u(t),t)$ contains the entire time-dependence as well as any
non-linearity of the generator. Since $s(u(t),t)$ depends on the not-yet-known
solution of the equation of motion, $u(t)$, Eq.~\eqref{eq:methods:inhom_eom} is
solved iteratively, until self-consistence is reached~\cite{ndong2010chebychev}.
To obtain the best possible convergence of the self-consistent loop, the
inhomogeneity should be as small as possible. This can be achieved by optimally
splitting the generator $G(t)$ into time-dependent and time-independent parts,
i.e., by taking the time-independent part at the mid-point of each time
interval, $G_0 \equiv G ((t_n + t_{n+1})/2)$ (as in the PWC approximation). The
time-dependent part is the `correction' to $G(t)$, i.e., $G_\text{td} (t) \equiv
G (t) - G_0$. Denoting the step in the self-consistent loop by $k$, $k \geq 1$,
Eq.~\eqref{eq:methods:inhom_eom} becomes
\begin{align}\label{eq:methods:iter_inhom_eom}
  \frac{\dd}{\dd t} u^{(k)}(t)
  =
  G_0 u^{(k)}(t) + s \big( u^{(k-1)}(t), t \big),
\end{align}
which requires a guess state $u^{(0)}(t)$ for the first step. The choice of
$u^{(0)}$ will be discussed below.

Provided the inhomogeneity $s$ can be written as a Taylor polynomial,
Eq.~\eqref{eq:methods:iter_inhom_eom} can be solved based on Duhamel's
principle. The latter links the solution of an inhomogeneous ODE to the
homogeneous solution $u_{\text{hom}}(t) = U(t) u_0$ by
\begin{equation}\label{eq:methods:duhamel}
  u_{\text{inhom}}(t) = u_{\text{hom}}(t) + \int_0^t U(t - \tau) s(\tau) \,\dd
  \tau.
\end{equation}
In our case, the homogeneous solution is simply given by
\[
  U(t)=\exp(G_0 t),
\]
since $G_0$ is time-independent. In order to obtain the required form for the
inhomogeneity, we first interpolate $s$ as an orthogonal polynomial of a given
order $M$~\cite{ndong2010chebychev,tal2012new,schaefer2017semi}. The `detour'
via orthogonal polynomial yields a global approximation of $s$ in the time
interval and thus much better convergence than directly evaluating $s$ in terms
of a Taylor series~\cite{RonnieReview94}. Note that this observation applies to
the comparison with any propagation method which converges only polynomially in
the time step and for which the error is distributed non-uniformly. This
includes in particular all propagation schemes constructed from the Taylor and
the Magnus expansion \cite{hochbruck_ostermann_2010}. Here, we choose Newton
polynomials for our global approximation because they open up the possibility to
increase $M$ on the fly, due to their recursive definition. For the time being,
we use constant $M$ and Chebyshev-Gauss-Lobatto (CGL) sampling points,
\begin{align}\label{eq:methods:cgl_points}
  \tau_j
  =
  \frac{\delta t}{2} \left(1 - \cos \Big( \frac{j-1}{M-1} \pi \Big)\right),
  && j = 1, \dots, M.
\end{align}
For a polynomial inhomogeneity $s$, the integral in
Eq.~\eqref{eq:methods:duhamel} can be solved analytically, yielding the formal
solution of Eq.~\eqref{eq:methods:iter_inhom_eom}, that is,
\begin{subequations}\label{eq:ITO}
  \begin{eqnarray}
    \label{eq:methods:u_sol_ts}
    u^{(k)} (t_n + \tau) = f_M (G_0, \tau) v_M^{(k)}
    + \sum_{m=0}^{M-1} \frac{\tau^m}{m!} v_m^{(k)},
  \end{eqnarray}
with
\begin{align}
     v_m^{(k)} = G_0 v_{m-1}^{(k)} + s_{m-1}^{(k)},
   \quad v_0^{(k)} = u^{(k)} (t_n)
\end{align}
and
\begin{align}
  f_M (z,t) =
  \begin{cases}
    \frac{1}{z^M} \bigg( \exp(zt) - \sum_{m=0}^{M-1} \frac{(zt)^m}{m!} \bigg)
    & z \neq 0 \\
  \frac{t^M}{M!} & z = 0.
  \end{cases}
\end{align}
\end{subequations}
The indices $k$ and $n$ in Eqs.~\eqref{eq:ITO} denote the current iteration and
time interval, respectively, with $\tau \in [0, \delta t]$. $s_m$ are the
coefficients of the Taylor-like polynomial obtained by interpolating the
inhomogeneity. The function $f_M$ will be computed using a spectral method,
analogously to evaluating $\exp(G_0 t)$ by expansion into Chebychev or Newton
polynomials \cite{tal1984accurate}. For more details see
Refs.~\cite{ndong2010chebychev,tal2012new,schaefer2017semi} as well as
Appendix~\ref{app:implementation}.

We now discuss how to choose the starting point $u^{(0)}$ of the iteration, cf.
Eq.~\eqref{eq:methods:iter_inhom_eom}. The proper choice of $u^{(0)}$ is of
high importance to the convergence as well as the stability of the iterative
process. We require knowledge of $u^{(0)}$ at the $M$ interpolation points
$t_n+\tau_j$, $j=1,\ldots, M$, in each time step, $[t_n,t_n+\delta t]$ to
evaluate $s_m$. In the following, we discuss three choices for the initial
guess.

(i)
Take $u^{(0)}(t)$ to be constant and equal to the value at the beginning of the
time step: $u^{(0)} (t_n+\tau_j) \coloneqq u(t_n)$. This is the zeroth order
approach, where we make use of the fact that the solution at the time $t_n$ has
already been obtained in the calculation for the previous time step. This
definition of the guess at each time grid point is the simplest possible
approach and requires no additional calculations. However, it turns out to be
the worst in terms of accuracy and leads to the largest number of iterations
required for convergence.

(ii)
Compute $u^{(0)}$ as solution to the homogeneous ODE~\cite{ndong2010chebychev}:
$u^{(0)} (t_n+\tau_j) \coloneqq u_{\text{hom}} (t_n+\tau_j)$ where
$u_{\text{hom}}$ is the solution to Eq.~\eqref{eq:methods:inhom_eom} when
setting $s \coloneqq 0$. It can be computed by one of the well-known quantum
propagators for time-independent generators~\cite{RonnieReview94}. In other
words, we use the solution obtained by a PWC approximation, and improve upon it
iteratively. Since this approach needs multiple matrix-vector operations to
determine $u^{(0)}$, it is more costly than option (i) regarding both CPU time
and memory. As it is a better guess, less iterations are required.
However, due to high numerical costs, it is still not the best option.

(iii)
Extrapolate the time-dependence of the full solution $u$ from the previous time
step \cite{schaefer2017semi} by evaluating Eq.~\eqref{eq:methods:u_sol_ts} for
$\tau_{j}$ shifted by $\delta t$, i.e., $u^{(0)} (t_n+\tau_j) \coloneqq
u_{t_{n-1}} (t_n+\tau_j)$. The idea is that, for a sufficiently smooth overall
solution $u_{t_{n-1}}$ in $[t_{n-1},t_{n}]$, it should provide a good initial
guess for the adjacent interval $[t_{n}, t_{n+1}]$. For small enough $\delta
t$,this choice of the guess should, on average, be the most accurate one. As for
the computational cost, no additional matrix-vector operations are necessary.
This can be seen by inspection of Eq.~\eqref{eq:methods:u_sol_ts}: All matrix
and vector components stay the same under extrapolation of $t=t_{n-1}+\tau_{j}$
to $t=t_{n}+\tau_{j}$; only the value of the parameter time changes. Hence, only
scalar coefficients have to be recalculated, which is negligible in terms of CPU
time. On the downside, however, the vector containing the coefficients of the
spectral approximation of $f_M$ as well as $v_m$ have to be stored for all
$\tau_j$ in order to be able to compute the extrapolation efficiently, making
this the most memory consuming method. Still, if the memory can be spared, it is
by far the most efficient of the three choices and recommended to be used.

\subsection{Quantum Optimal Control with Iterative Time Ordering}
\label{subsec:methods:itoqoct}
Quantum optimal control theory (OCT) provides methods to compute
controls, i.e., external fields $\{ \pulse_k \}$ interacting with the
quantum system, that steer the system's dynamics in a desired
way~\cite{GlaserEPJD15}. A cost functional, defined here for a single
field $\pulse(t)$,
\begin{align} \label{eq:methods:J}
  J\big[\pulse \big]
  &=
  J_{T}\big[\ket{\psi(T)} \big]
  + \int_{0}^{T} g\big[\pulse(t), \ket{\psi(t)}, t\big] \,\dd t,
\end{align}
has to be minimized. $J_T$ is the final-time functional, indicating the figure
of merit, and
\begin{align}
  g\left[ \pulse(t), \ket{\psi(t)}, t\right] =
    g_a \left[ \pulse (t), t \right] +
    g_b \left[ \ket{\psi (t)}, t \right] \nonumber
\end{align}
encodes the intermediate-time costs. Here, we use Krotov's method
\cite{Konnov.AutomRemContr.60.1427, SklarzPRA02, Palao.PRA.68.062308,
Reich.JCP.136.104103}, a gradient-based sequential optimization algortihm with
guaranteed monotonic convergence. With the choice
\begin{align*}
  g_a \left[ \pulse(t), t \right] =
  \frac{\lambda_a}{S(t)} \left( \pulse(t) - \pulse_{\mathrm{ref}}(t) \right)^2,
\end{align*}
the extremum condition on the functional $J$ with respect to $\pulse$ yields the
following equation for the `new' field~\cite{Palao.PRA.68.062308}
\begin{subequations}\label{eq:methods:stcost}
\begin{align}\label{eq:methods:eps_update}
  \pulse^{(i+1)}(t) &= \pulse_{\mathrm{ref}}(t) \, + \\
    &\hphantom{=.} \frac{S(t)}{\lambda_a} \ip 
    \Braket{\chi^{(i)}(t) \Bigg\vert
    \frac{\partial \op{H}}{\partial \pulse}
    \bigg\vert_{\substack{\pulse^{(i+1)} \\ \psi^{(i+1)}}}
    \Bigg\vert \psi^{(i+1)}(t)}. \nonumber
\end{align}
Here, $i$ denotes the iteration step in the optimization process.
The subscripts of $\partial \op{H}/\partial \pulse$ indicate that the
derivative may depend on both the state and on the control field and
has to be evaluated at the current iteration.
The reference field $\pulse_{\mathrm{ref}}$ is often chosen as
$\pulse^{(i)}$, i.e., the `old' field, to
yield a direct update formula for $\Delta \pulse^{(i)} = \pulse^{(i+1)}
- \pulse^{(i)}$.
The equation of motion together with its initial condition,
\begin{align}
  \frac{\partial}{\partial t} &\ket{\psi^{(i+1)}(t)} = - \frac{\ci}{\hbar}
  \op{H}(\pulse^{(i+1)}) \, \ket{\psi^{(i+1)}(t)}, \label{eq:methods:st_se} \\
  &\ket{\psi^{(i+1)}(0)} = \ket{\psi_0}, \label{eq:methods:st_initial}
\end{align}
are an input to the algorithm, whereas the extremum condition on $J$ with
respect to the state yields
\begin{align}
  \frac{\partial}{\partial t} &\ket{\chi^{(i)}(t)} = - \frac{\ci}{\hbar}
  \op{H}^\dagger(\pulse^{(i)}) \, \ket{\chi^{(i)}(t)} + \nabla_{\bra{\psi}} \,
  g_b \big\vert_{\psi^{(i)}(t)}, \label{eq:methods:cost_se} \\
  &\ket{\chi^{(i)}(T)} = - \nabla_{\bra{\psi}} \, J_T \big\vert_{\psi^{(i)}(T)}.
  \label{eq:methods:cost_initial}
\end{align}
\end{subequations}
Equation~\eqref{eq:methods:eps_update} assumes that $J_T$ depends at most
quadratically on the state $\ket{\psi(T)}$ which is the case e.g.\ for
expectation values. For more complicated dependencies of $J_T$ on the state,
a second term appears in the rhs of Eq.~\eqref{eq:methods:eps_update}
\cite{Reich.JCP.136.104103}.

Equations~\eqref{eq:methods:stcost} are a set of non-linear coupled equations
whose numerical evaluation is not trivial. A common approximate solution is
based on the discretization of the time grid~\cite{Palao.PRA.68.062308}. It
consists in using the known state $\ket{\psi^{(i+1)}(t_n)}$ instead of the
required but unknown state $\ket{\psi^{(i+1)}(t_{n+1})}$ to obtain the updated
pulse at the next time step, $\pulse^{(i+1)}(t_{n+1})$. With the ITO
propagator, we no longer rely on such an approximation to calculate the
(effectively) non-linear propagation. However, when solving the equations of
motion in Krotov's method with the ITO propagator, two self-consistent loops
have to be combined. The control loop counts the updates of the field and is
indexed by the superscript $(i)$, cf. Eq.~\eqref{eq:methods:eps_update}. Within
one step of the control loop, we employ a time discretization, i.e., we evaluate
Eq.~\eqref{eq:methods:eps_update} for $0<t_n\le T$. This loop over $n$ is
a regular loop, not involving any self-consistency. For a given control
iteration $i$ and time step $t_n$, the ITO loop with index $k$ improves upon an
initial guess to determine the true state, cf. Eq.~\eqref{eq:methods:u_sol_ts}.
Since Eq.~\eqref{eq:methods:eps_update} requires knowledge of the `new' state,
at control iteration $i+1$, the loop over $(i)$ has to be the outermost loop.
However, determination of the `new' state $\ket{\psi^{(i+1,k)}(t_{n+1})}$ within
the innermost (ITO) loop over $k$, requires knowledge of field
$\pulse^{(i+1,k)}(t_{n+1})$ in order to evaluate the Hamiltonian. In fact, it
does so not only at the sampling points of the global time grid $t_n$ but also
within each time interval $\delta t$, i.e., for all $t_n+\tau_j$. In other
words, the inhomogeneity now involves two unknowns that must be determined
self-consistently -- the field $\pulse^{(i+1,k)}(t_n+\tau_j)$ and the state
$\ket{\psi^{(i+1,k)}(t_n+\tau_j)}$.

Our approach to resolve this mutual dependence consists in updating the field
$\pulse^{(i+1,k)}(t_n+\tau_j)$ alongside the state
$\ket{\psi^{(i+1,k)}(t_n+\tau_j)}$ within the ITO loop. In more detail, the ITO
loop, cf. Eq.~\eqref{eq:methods:u_sol_ts}, is initialized by choosing an initial
guess for the state, $\ket{\psi^{(i+1,k=0)}(t_n+\tau_j)}$, just as in the
original ITO propagator. Unlike in this case, where the field is assumed to be
known, $\pulse^{(i+1,k=0)}(t_n+\tau_j)$ is now calculated from
$\ket{\psi^{(i+1,k=0)}(t_n+\tau_j)}$, using Eq.~\eqref{eq:methods:eps_update}.
This is the input for the actual ITO loop that calculates
$\ket{\psi^{(i+1,k=1)}(t_n+\tau_j)}$ from Eq.~\eqref{eq:methods:u_sol_ts}. The
updated state $\ket{\psi^{(i+1,k=1)}(t_n+\tau_j)}$, in turn, yields
$\pulse^{(i+1,k=1)}(t_n+\tau_j)$ which is the input for the next step of the ITO
loop, resulting in $\ket{\psi^{(i+1,k=2)}(t_n+\tau_j)}$ and so forth. This
procedure of conjointly updating the field and the state within the time
interval $[t_n,t_{n+1}]$ is repeated until self-consistency is reached. The
algorithm then advances to the next time step $t_{n+2}$.

Our ansatz can be motivated as follows. Recalling the Krotov update
equation~\eqref{eq:methods:eps_update}, the underlying problem is analogous to
the one treated in the derivation of the propagation algorithm: What prevents
solution of Eq.~\eqref{eq:methods:eps_update} in closed form is the (implicit)
presence of the lhs of the equation also in the rhs, since the state
$ \ket{\psi^{(i+1)}(t)}$ is propagated under the updated pulse
$\pulse^{(i+1)}(t)$. Similarly, in case of the ITO propagator, the solution
$\Ket{\psi(t)}$ is present in the inhomogeneity. One can thus think of treating
the non-linearity of the control equations~\eqref{eq:methods:stcost} as an
inhomogeneity that needs to be determined self-consistently. The most balanced
way to determine the interdependent field and state is to do it in the
interleaved fashion described above.

One may wonder whether self-consistent determination of the field is really
necessary. A simple alternative would be to calculate
$\pulse^{(i+1,k=0)}(t_n+\tau_j)$ only once at the initialization stage of the
ITO loop (from the guess for the state) and omit updating it. This turns out to
not work at all. In other words, the estimate $\pulse^{(i+1,k=0)}(t_n+\tau_j)$
is not sufficiently close to the true field, and without knowledge of the true
field, convergence of the state to the correct one is not possible either.

The ITO approach to Krotov's method in quantum control is a quite natural one,
since it is closer to the time-continuous original, derived for classical
mechanics applications~\cite{Konnov.AutomRemContr.60.1427}. In quantum control,
the update equation was discretized in order to numerically solve
it~\cite{SklarzPRA02,Palao.PRA.68.062308}. The discretization is still required
for the global time interval $[0, T]$, in order to get a sequential update
scheme that evaluates the `new' field at each time $t_n$. However, we no longer
need an approximation for the time interval $[t_n,t_{n+1}]$, i.e., on the small
scale of a single time step $\delta t$. Instead of replacing the actually
required state by the known one from the previous time
step~\cite{Palao.PRA.68.062308}, the update
formula~\eqref{eq:methods:eps_update} is now solved in its continuous form by
iteratively converging it. Just as the ITO propagator, Krotov's method is now
solved semi-globally with respect to the time steps.

\section{Numerical Benchmark: Harmonic Oscillator}
\label{sec:bench}
In this section we analyze both the propagator alone as well as its application
in OCT with respect to their numerical performance. Of particular interest is
the efficiency, i.e., the quality of the solution per numerical effort.

\subsection{Benchmarking the ITO Propagator}
\label{sec:bench:itoprop}
In order to determine the accuracy of a propagation, we require a
physical system with external driving that has an exact analytical solution.
A system which matches this requirement without being numerically trivial is
the linearly driven harmonic oscillator (HO). The Hamiltonian reads
\begin{equation}\label{eq:bench:H_QHO}
  \op{H}_\text{HO}(t) = \frac{\op{p}^2}{2 m} + \frac{1}{2} m \omega^2 \op{x}^2
    + \pulse(t) \op{x},
\end{equation}
where $\op{x}$ and $\op{p}$ denote the position and momentum operators,
respectively, and $m$ and $\omega$ are mass and frequency of the oscillator.
We assume a driving field of the form
\begin{equation}\label{eq:bench:HO_pulse}
  \pulse(t) = \pulse_0 \sin^2 \left(\pi t/T \right)
    \cos \left( \omega_L t \right).
\end{equation}
In addition to the time-dependence of the envelope, there are fast
oscillations with frequency $\omega_L$. This parameter allows us to
control the strength of the time-dependence of the calculations.

The analytical solution of the driven HO~\eqref{eq:bench:H_QHO} is known for
the case that the system is initially prepared in an eigenstate of the
undriven HO such as the ground state $\Psi_0(x) = \braket{x \vert
0}$~\cite{SalaminJPhysA1995}.
One first has to compute
\begin{equation}\label{eq:bench:HO_z}
  z(t) = - \exp \left( \ci \omega t \right)
    \int_0^t \pulse(\tau) \exp \left( - \ci \omega \tau \right) \,\dd\tau,
\end{equation}
the expectation values for position and momentum are then associated with the
imaginary and real parts
\begin{align}
  \braket{x(t)} = \ip ( z(t) ),
  && \braket{p(t)} = \rp ( z(t) ). \nonumber
\end{align}
For our choice of pulse, Eq.~\eqref{eq:bench:HO_pulse}, the integral
in~\eqref{eq:bench:HO_z} can be solved analytically.

We will analyze the numerical stability and efficiency of the ITO
propagator in terms of two parameters, the size of the time step,
$\delta t$, and the expansion order of the inhomogeneity, $M$.
For the ITO propagator, $\delta t$ has to be chosen carefully, as a bad choice
compromises the convergence behavior. $M$ corresponds to the number of
sampling points $\tau_j$ in each local time grid, i.e., within the
interval $[t, t+\delta t]$. Since $M$ determines the accuracy of the
approximation of the inhomogeneity, it affects how well the solution
is improved in each iteration. As a consequence, it potentially has a
large impact on the required number of iterations. Hence, a good
choice of $M$ is imperative both for efficiency and stability of
the propagator.

\begin{figure}
  \centering
  \includegraphics{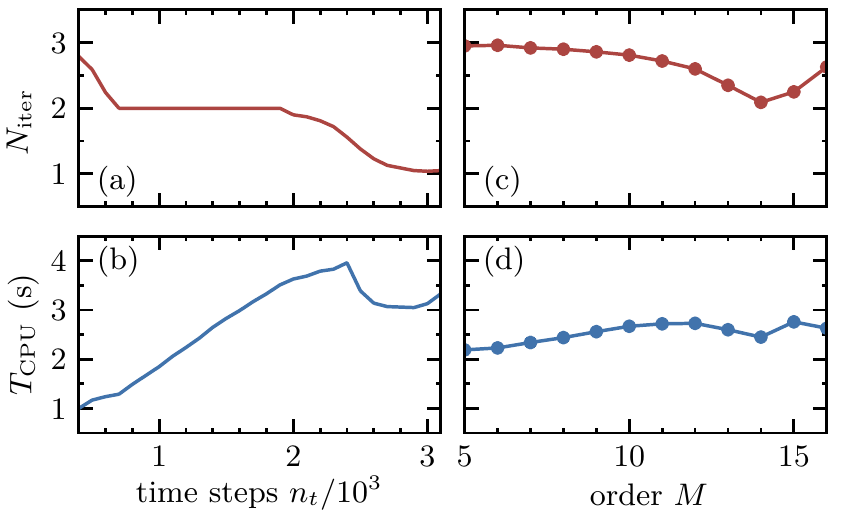}
  \caption[$N_{iter}$ and $T_{CPU}$ vs. $n_t$ and $M$ for ITO]{%
    Analysis of the ITO propagator's numerical efficiency: Mean number of
    iterations per time step $N_{\mathrm{iter}}$ and elapsed CPU time
    $T_\text{CPU}$ vs.\ the total number of time steps $n_t$ (a, b) and the
    expansion order $M$ (c, d). The system parameters (cf.
    Eqs.~\eqref{eq:bench:H_QHO}-\eqref{eq:bench:HO_pulse}) are $m=\omega=1$,
    $\pulse_0=10^{-3}$, $\omega_L=5$, $T=100$ with $M = 8$ (a, b) and $n_t=900$
    (c, d); the desired accuracy is $\epsilon = 10^{-12}$.
  \label{fig:bench:ito_nt_m}}
\end{figure}
We analyze the numerical effort as a function of both parameters in
Fig.~\ref{fig:bench:ito_nt_m}. The upper panel shows the number of iterations
$N_{\mathrm{iter}}$ required to reach an accuracy of $10^{-12}$ as a function of
the number of time steps $n_t$ (which is inversely proportional to $\delta t$
for fixed $T$)
and the order $M$ of the inhomogeneity. The lower panel displays additionally
the elapsed CPU time $T_{\mathrm{CPU}}$ for that computation~\footnote{The
 computer used for all computations is an Intel Core i7-5930 @ 3.50\,GHz system
with 32\,GB RAM and a 64-bit Linux OS.}. A direct correlation between the
required number of iterations and the CPU time is observed: In
Fig.~\ref{fig:bench:ito_nt_m}(b), two sharp bends occur -- the first one at the
point where $2$ iterations begin to suffice to reach convergence, at roughly
$n_t = 700$; prior to that, on average, $2$ to $3$ iterations were required.
Between $n_t=700$ and $1900$, there is a plateau in the required number of
iterations, since the increase of $n_t$, i.e.\ decrease of $\delta t$, did not
suffice to reach convergence within less iterations, although the accuracy (per
iteration) was increased. During this plateau, the increased number of time
steps thus only leads to a larger numerical effort, as can be seen in the CPU
time. The second bend is the larger one, at $n_t=1900$, where the propagator
first was able to reach convergence within one single iteration, for some of the
time steps, leading to $N_{\mathrm{iter}}$ between $1$ and $2$. Further
decrease of this quantity due to decrease of $\delta t$ rapidly improved the
CPU time, until -- again -- a plateau is reached at $N_{\mathrm{iter}} = 1$. At
this point, a further decrease of $\delta t$ is not reasonable, since it can
only increase the numerical effort without any gain, because less than one
iteration is of course not possible.

A similar relation between $T_\text{CPU}$ and $N_{\mathrm{iter}}$ can be
observed in Fig.~\ref{fig:bench:ito_nt_m}(c, d). For constant $\delta t$, an
increase in $M$ leads to higher CPU times up to the point where the number of
iterations is noticeably reduced below $3$. The effort decreases then, until
$N_{\mathrm{iter}}=2$ is reached. Interestingly, $N_{\mathrm{iter}}$ increases
again when further increasing $M$. This is counter-intuitive, since a higher
order should decrease the number of iterations. However, for too high orders $M$
numerical instabilities occur in the interpolation of the inhomogeneity in
orthogonal polynomials. In fact, for $M > 16$ the procedure does not converge at
all. Failure to reach convergence happens, of course, also for too small $M$
since then the inhomogeneity is not represented accurately enough. Between
these two limits, the curve in Fig.~\ref{fig:bench:ito_nt_m}(d) shows only
a weak dependence on the order $M$, indicating a surprisingly low impact of this
parameter on the efficiency. This can be explained as follows: Larger $M$
increases the cost of each iteration while, at the same time, decreasing the
average number of iterations, $N_{\mathrm{iter}}$. The two effects approximately
cancel out in the end, leading to an almost constant dependence of
$T_\text{CPU}$ on $M$. In conclusion, $M$ should be chosen carefully in
accordance with the time step size, with $M = 8$ representing a good starting
point.

We now address the question of how iterative time ordering compares to
the PWC approximation. For the latter, we employ the Chebyshev
propagator \cite{tal1984accurate} where, for a given $\delta t$, the number of
coefficients $n_\text{c}$ required in order to reach machine precision
are calculated. Hence, the time-independent problem is solved with
maximal precision and the only inaccuracy is due to the PWC
approximation. In other words, there is only one free parameter, which
also determines precision -- $\delta t$ or the number of time steps $n_t$.

\begin{figure}
  \centering
  \includegraphics{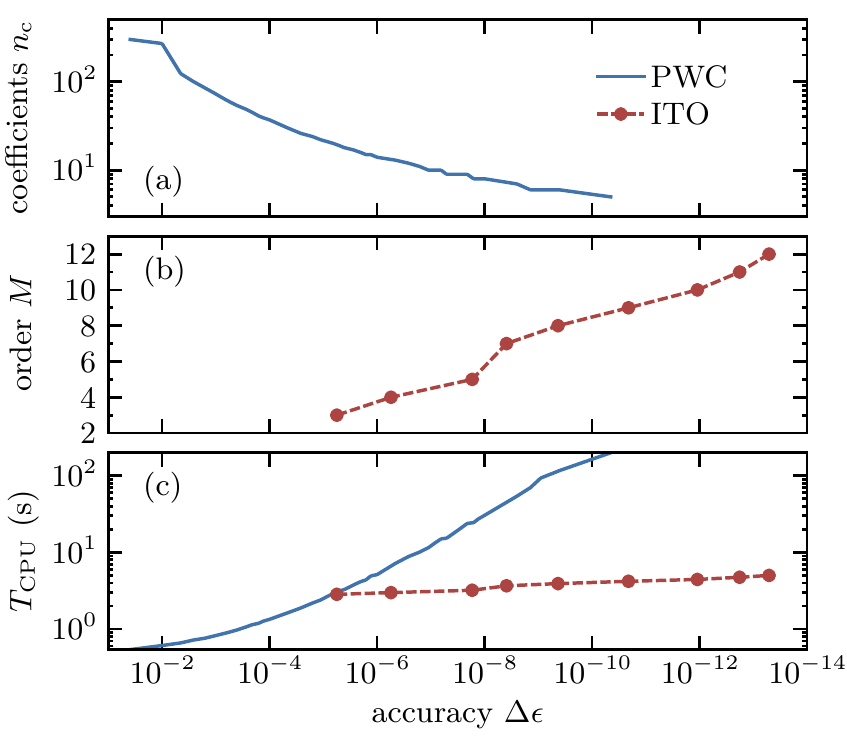}
  \caption[Comparison of ITO vs. PWC (Chebyshev) propagators]{%
    Comparison of iterative time ordering and the PWC approximation using the
    Chebyshev propagator: (a) Number of Chebyshev coefficients $n_\text{c}$, (b)
    ITO expansion order $M$, and (c) elapsed CPU time $T_\text{CPU}$ vs.\
    desired accuracy $\Delta \epsilon$. The system parameters are as in
    Fig.~\ref{fig:bench:ito_nt_m}, except for $\omega_L = 1.001$, corresponding
    to near-resonant driving, and $T=1000$.
  \label{fig:bench:ito_cheby}}
\end{figure}
Figure~\ref{fig:bench:ito_cheby} compares the CPU time required to reach
a certain accuracy for the two propagation approaches. The accuracy corresponds
to the maximal deviation of the numerical solutions from the analytical one.
For the Chebyshev propagator and the PWC approach, we continuously increased
$n_{t}$, respectively decreased $\delta t$, in order to continuously improve the
quality of the PWC approximation. Figure~\ref{fig:bench:ito_cheby}(a) shows the
number of required Chebyshev coefficients $n_{\text{c}}$ within each time step
$\delta t$. We observe a strong decrease of $n_\text{c}$ (as $\delta t$
decreases from left to right) from about $300$ to $5$ in the end. For the ITO
propagator, the number of time steps was set constant to $n_t = 4000$, and only
the order $M$ was varied, between $3$ and $12$, as shown in
Fig.~\ref{fig:bench:ito_cheby}(b). This leads to accuracies between $6 \cdot
10^{-5}$ and $5 \cdot 10^{-14}$. Larger inaccuracies could not be realized with
the ITO propagator. The lower limit is close to, but above machine precision and
is due to accumulation of errors. Accumulation of errors is of course also --
and especially -- a problem for the PWC approach, where up to $10$ million time
steps were required to achieve the higher accuracies. The smallest error that
can realized in the PWC approach amounts to about $\Delta \epsilon \sim
10^{-11}$; further decreasing $\delta t$ does not push the error to below this
value. In contrast, the ITO propagator avoids the accumulation of errors and
allows for realizing errors much closer to machine precision over the complete
propagation time interval.

The CPU time increases almost linearly with the accuracy for the PWC
propagations in the double-logarithmic plot as well, corresponding to an
approximately cubic dependence. For the ITO propagator, the increase in CPU time
with the accuracy is significantly weaker. The required CPU time varied from
$\SI{2.84}{s}$ to $\SI{5.01}{s}$ for ITO and from $\SI{0.53}{s}$ to
$\SI{199}{s}$ for the PWC approximation. An interesting point is at about
$\Delta\epsilon=10^{-5}$, where the two curves nearly intersect. If an accuracy
higher than this value is required, the ITO propagator easily outperforms the
PWC approach; the higher the accuracy, the more obvious this is. For lower
accuracies, however, the PWC approach remains the best choice, being both stable
and very fast. It should be noted that this threshold of $\Delta\epsilon$
cannot be generalized, since it depends on the strength of the time dependence
of the problem at hand. While no rigorous measure to quantify the strength of
the time dependence has, to the best of our knowledge, so far been brought
forward, it is increased (decreased) in our example in
Fig.~\ref{fig:bench:ito_cheby} by increasing (decreasing) $\omega_{L}$ or
$\pulse_{0}$ in Eq.~\eqref{eq:bench:HO_pulse}. As a consequence, the threshold
of $\Delta\epsilon$ decreases (increases) accordingly.

\subsection{Benchmarking Krotov's Method for Quantum Optimal Control with
Iterative Time Ordering}
\label{sec:bench:itoqoct}
\begin{figure}
  \centering
  \includegraphics{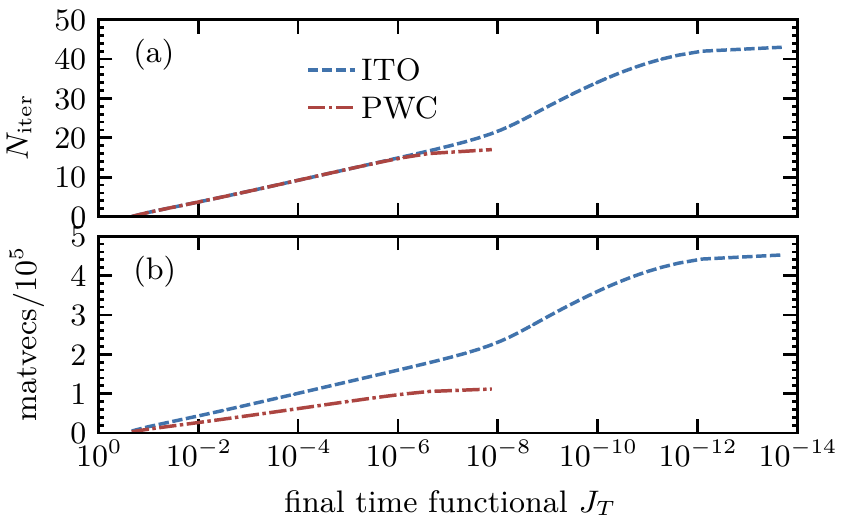}
  \caption[PWC/ITO Krotov benchmark HO]{%
    Comparison of the ITO (dashed) and PWC (dash-dotted) propagators when used
    in Krotov's method: (a) Number of iterations $N_{\text{iter}}$ and (b)
    number of matrix-vector operations required to reach a certain value of the
    functional $J_T$. The considered system is the HO as in
    Eq.~\eqref{eq:bench:H_OCT} where the frequency is controlled ($T = 2$,
    $\delta t = 0.01$, $M = 5$).
  \label{fig:bench:krito_ho}}
\end{figure}
We now determine the numerical effort required to reach a certain quality of the
optimization result. The system to be controlled is once again the HO, this
time described by a slightly different Hamiltonian,
\begin{align}\label{eq:bench:H_OCT}
  \op{H}_\text{freq}(t) = \frac{\op{p}^2}{2} + \frac{1}{2} \op{x}^2 \pulse (t),
\end{align}
i.e., we set $\omega=m=1$. Equation~\eqref{eq:bench:H_OCT} represents an example
which is controllable, in contrast to Eq.~\eqref{eq:bench:H_QHO}, which is
uncontrollable due to the equidistant energy levels. The control function
$\pulse (t)$ effectively represents the frequency (or rather its square) of the
harmonic potential which alters the system's eigenstates. We begin in the ground
state $\ket{\psi_0}$ of the HO for $\omega=1$ and seek to transfer it into the
ground state of a HO with $\omega=1/2$, by varying the harmonic potential's
frequency via $\pulse (t)$. The effort is measured in terms of number of
iterations $N_{\text{iter}}$ and performed matrix-vector operations to reach
a given fidelity or final time functional value,
\begin{equation}\label{eq:state2state}
  J_T [\psi (T)] = 1 - | \braket{\psi(T) | \psi_\text{target}} |^2.
\end{equation}
$J_T$ measures the difference between the final state $\ket{\psi(T)}$ and the
target state $\ket{\psi_{\text{target}}}$.

The results of the benchmark are shown in Fig.~\ref{fig:bench:krito_ho}. The
improvement per control iteration is very similar for both methods up until
around $J_T \approx 10^{-6}$ with ITO incurring a higher numerical cost. When
continuing the optimization towards even more accurate controls, the PWC method
could decrease $J_T$ further to around $10^{-8}$, whereas ITO reached a value of
nearly $10^{-14}$. Again, being a more costly method than PWC, ITO requires
a larger number of operations for an equal amount of iterations, cf.
Fig.~\ref{fig:bench:krito_ho}(b). In contrast to the PWC approximation with
which Krotov's method breaks down at around $J_T=10^{-8}$ since the errors in
the time propagation do not allow for further improvements of the control,
Krotov's method with the ITO propagator can reach the control target with
essentially arbitrary (i.e., close to machine precision) accuracy.

The solutions are found to be identical up to slight variations which are due to
the higher accuracy of the ITO propagator and account for the difference in the
fidelities beyond $J_T \approx 10^{-6}$. We conclude that, for the HO, the
combination of Krotov's method with iterative time ordering allows for more
accurate control solutions. However, this will turn out to be not a general
feature, as we show below. The higher accuracy comes at the price of a larger
numerical effort which increases linearly for exponentially smaller errors.

\section{Pythagorean Control of a Superconducting Qudit under Strong Driving}
\label{sec:pc}
A more realistic application for the previously introduced propagator occurs in
the dynamics of superconducting qubits, where driving fields are often rapidly
oscillating in time. Here, we consider a superconducting qubit under Pythagorean
control \cite{Suchowski.PRA.84.013414, Svetitski.NatComm.5.5617}.

\subsection{Model}
\label{subsec:pc:model}
Quantized superconducting circuits can encode information in their lowest few
energy eigenstates, since the dynamics can be confined experimentally to these
relevant levels \cite{npjQuantumInf.3.2}. For the lowest eigenstates, the
`qubit' can be modeled by an anharmonic ladder, not necessarily ending after two
(qubit) levels. Therefore, we use the term qudit in the following. The
Hamiltonian for a driven $N$-level qudit reads
\begin{align} \label{eq:pc:ham}
  \op{H}(t)
  =
  \op{H}_{0} + \op{H}_{1}(t)
\end{align}
with drift Hamiltonian ($\hbar = 1$)
\begin{align}
  \op{H}_{0}
  =
  \sum_{n=0}^{N-1} \epsilon_{n} \Ket{n}\Bra{n},
  &&
  \epsilon_{n} = n \omega_{0} - \frac{\beta}{2} n(n-1), \notag
\end{align}
where $\epsilon_{n}$ is the eigenenergy for eigenstate $\Ket{n}$ of
$\op{H}_{0}$. The parameter $\omega_{0}$ defines the base energy difference
between adjacent levels of the qudit and $\beta$ determines its anharmonicity.
The control Hamiltonian is given by
\begin{align}
  \op{H}_{1}(t)
  =
  \sum_{n=0}^{N-2} \sqrt{n+1} \Big[%
    \Ket{n}\Bra{n+1} + \text{H.c.}
    \Big] \pulse(t), \notag
\end{align}
where $\pulse(t)$ is the external control field and $\text{H.c.}$ denotes the
Hermitian conjugate.

As experimentally demonstrated in Ref.~\cite{Svetitski.NatComm.5.5617},
population inversion between non-adjacent levels of the four-level system can be
realized using so called Pythagorean couplings \cite{Suchowski.PRA.84.013414}.
The corresponding external field is given by \cite{Svetitski.NatComm.5.5617}
\begin{align} \label{eq:pc:field}
  \pulse(t)
  =
    \frac{V_{01}}{\sqrt{1}} \cos\left(\omega_{01} t\right)
  + \frac{V_{12}}{\sqrt{2}} \cos\left(\omega_{12} t\right)
  + \frac{V_{23}}{\sqrt{3}} \cos\left(\omega_{23} t\right),
\end{align}
with $\omega_{ij} = \epsilon_{j} - \epsilon_{i}$ and $V_{ij}$ the driving
strength of transition $\omega_{ij}$. Transforming Eq.~\eqref{eq:pc:ham} into
the interaction picture yields (see Appendix~\ref{app:derivation} for details)
\begin{align} \label{eq:pc:ham_int}
  \op{H}_{\text{int}}(t)
  &=
  \frac{1}{2}
    \begin{pmatrix}
      0      & V_{01} & 0      & 0      & 0      & \dots  & 0 \\
      V_{01} & 0      & V_{12} & 0      & 0      & \dots  & 0 \\
      0      & V_{12} & 0      & V_{23} & 0      & \dots  & 0 \\
      0      & 0      & V_{23} & 0      & 0      & \dots  & 0 \\
      0      & 0      & 0      & 0      & 0      & \dots  & 0 \\
      \vdots & \vdots & \vdots & \vdots & \vdots & \ddots & 0 \\
      0      & 0      & 0      & 0      & 0      & \dots  & 0 \\
    \end{pmatrix}
    + \op{H}_{\text{rot}}(t)
  \notag
  \\
  &\equiv
  \op{H}_{\text{inf}} + \op{H}_{\text{rot}}(t).
\end{align}
The first, time-independent term $\op{H}_{\text{inf}}$ matches exactly the
requirements for Pythagorean control \cite{Suchowski.PRA.84.013414}, while the
second, time-dependent term $\op{H}_{\text{rot}}(t)$ contains co- and
counter-rotating terms due to the rotating frame, cf.
Appendix~\ref{app:derivation}. We neglect this term for a moment and assume
Hamiltonian~\eqref{eq:pc:ham_int} to be entirely given by $\op{H}_{\text{inf}}$.
In this case, the control scheme has been derived analytically
\cite{Suchowski.PRA.84.013414}. In order to achieve population inversion between
$\Ket{0}$ and $\Ket{2}$ (which we will consider in the following), the driving
strengths $V_{ij}$ must be scaled by a primitive Pythagorean triple
\cite{Suchowski.PRA.84.013414},
\begin{align} \label{pc:eq:triple_scaling}
  \left(V_{01}, V_{12}, V_{23}\right)
  =
  \Omega_{\text{rabi}}
  \left( \frac{p^{2}+q^{2}}{2}, p q, \frac{p^{2}-q^{2}}{2} \right),
\end{align}
with $p, q$ odd integers. For practical purposes, since the total magnitude of
all $V_{ij}$ can be scaled by an experimentally accessible parameter (the Rabi
frequency $\Omega_{\text{rabi}}$), we allow $p,q$ to take non-integer values.

Unfortunately, the additional term $\op{H}_{\text{rot}}(t)$ in
Eq.~\eqref{eq:pc:ham_int} vanishes only in the unphysical limit of infinite
anharmonicity $\beta$. We will analyze deviations below. To account for
dissipative effects due to the interaction of the superconducting circuit with
its environment, we employ a Lindblad master equation \cite{Breuer.book}
\begin{equation}
  \begin{aligned}
    \im \frac{\partial}{\partial t} \op{\rho}(t)
    &=
    \big[\op{H}(t), \op{\rho}(t)\big]
    +
    \lop{L}_{D}\left[\op{\rho}(t)\right],
    \\
    \lop{L}_{D}\left[\op{\rho}(t)\right]
    &=
    \im \sum_{k=1}^2 \Big(%
        \op{L}_{k} \op{\rho}(t) \op{L}_{k}^{\dagger}
        - \frac{1}{2} \big\{%
          \op{L}_{k}^{\dagger} \op{L}_{k}, \op{\rho}(t)
        \big\}
      \Big), \notag
  \end{aligned}
\end{equation}
with Lindblad operators \cite{Reich.SciRep.5.12430}
\begin{align}
  \op{L}_{1}
  =
  \sum_{n=0}^{N-2} \sqrt{\frac{n+1}{T_{1}}} \Ket{n} \Bra{n+1},
  &&
  \op{L}_{2}
  =
  \sum_{n=1}^{N-1} \sqrt{\frac{2 n^{2}}{T_{2}^{*}}} \Ket{n} \Bra{n}, \notag
\end{align}
where $T_{1}$ is the population relaxation time and $T_{2}^{*}$ the pure
dephasing time. The parameters of Tab.~\ref{tab:pc:parameters} are used in the
following and the qudit ladder is truncated at $N=10$ levels, which was observed
to suffice.

\begin{table}
  \centering
  \caption{%
    The parameters for the Pythagorean controlled qudit, taken from Ref.
    \cite{Svetitski.NatComm.5.5617}.
  }
  \begin{tabular*}{\linewidth}{c@{\extracolsep{\fill}}cc}
    \hline
    qudit frequency  & $\omega_{0}/2\pi$           & $6.73$\,GHz \\
    anharmonicity    & $\beta/2\pi$                & $0.12$\,GHz \\
    relaxation time  & $T_{1}$                     & $230$\,ns   \\
    dephasing time   & $T_{2}$                     & $120$\,ns   \\
    Rabi freqeuncy   & $\Omega_{\text{rabi}}/2\pi$ & $47.6$\,MHz \\
    field parameters & $p,q$                       & $0.86$      \\
    \hline
  \end{tabular*}
  \label{tab:pc:parameters}
\end{table}

\begin{figure}
  \centering
  \includegraphics{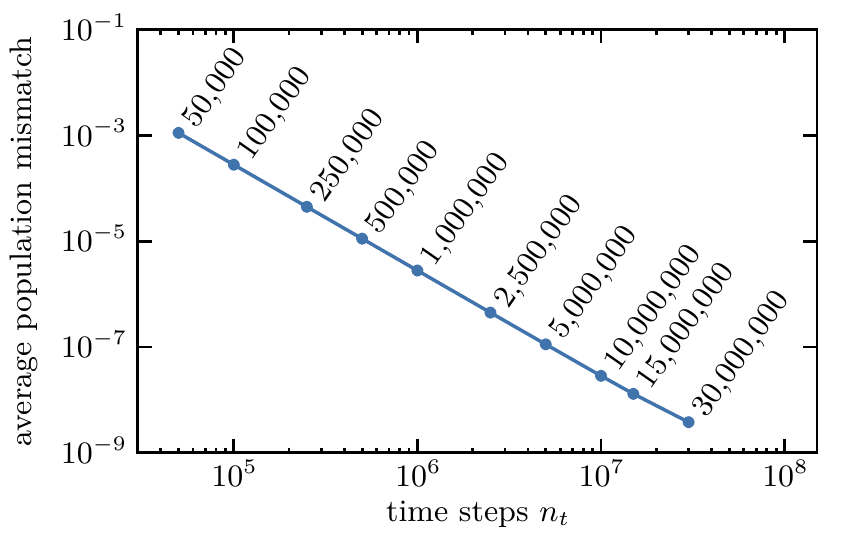}
  \caption{%
    Average population mismatch when comparing propagations using the PWC method
    against a single propagation using ITO\@. The total propagation time is
    $T=150$\,ns. Shown is the time-averaged maximal population mismatch, cf.
    Eq.~\eqref{eq:pc:pop_diff}. All simulations using the PWC propagator are
    compared against the same ITO propagation with $n_{t}=50{,}000$ constant
    time steps and $M=12$.
  }
  \label{fig:diff_ito_pwc_pc}
\end{figure}

\subsection{Numerical Considerations for the ITO Propagator}
\label{subsec:pc:itoprop}
\begin{figure}
  \centering
  \includegraphics{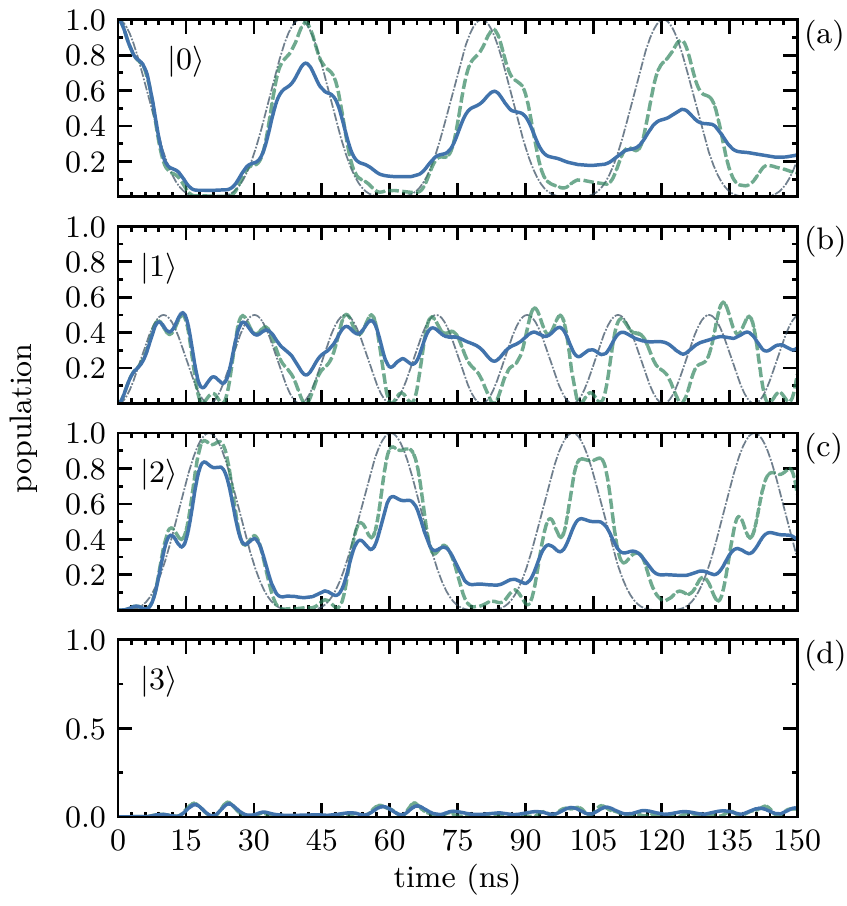}
  \caption{%
    Population dynamics in the qudit subspace $\left\{\Ket{0}, \Ket{1}, \Ket{2},
    \Ket{3}\right\}$. The dash-dotted lines represent the ideal, non-dissipative
    case using only $\op{H}_{\text{inf}}$, cf. Eq.~\eqref{eq:pc:ham_int}. The
    dashed (solid) lines show the non-dissipative (dissipative) population
    dynamics under Hamiltonian~\eqref{eq:pc:ham}. The dissipative dynamics
    corresponds to the convergence analysis in Fig.~\ref{fig:diff_ito_pwc_pc}.
  }
  \label{fig:pop_dyn}
\end{figure}

Figure~\ref{fig:diff_ito_pwc_pc} proves that the ITO propagator also works for
more complex systems than the HO; here for the qudit under Pythagorean control.
The respective population dynamics (obtained with ITO) is depicted in
Fig.~\ref{fig:pop_dyn}. Since an analytical solution for the dynamics is not
available in this case, we compare propagations using a PWC propagator (with an
increasing number of time steps) to a single propagation using ITO\@. Here, we
assume that any PWC propagator converges to an accurate time-ordered solution
provided the time discretization is sufficiently fine. We quantify the mismatch
between both simulations by taking the time-average of the function
\begin{align} \label{eq:pc:pop_diff}
  P_{\text{mis}}(t)
  =
  \max_{n=0,\dots,N-1} \left|%
      P_{n}^{\text{PWC}}(t) - P_{n}^{\text{ITO}}(t)
    \right|
\end{align}
with $P_{n}^{\text{PWC}}(t)$, $P_{n}^{\text{ITO}}(t)$ the populations in the
$n$th level obtained with PWC or ITO propagation, respectively. As can be seen
in Fig.~\ref{fig:diff_ito_pwc_pc}, the mismatch of the population dynamics
between PWC and ITO propagation decreases linearly with increasing number of PWC
time steps $n_{t}$ in a double logarithmic plot. Since we assume increasing
accuracy of the PWC propagation for increasing $n_{t}$,
Fig.~\ref{fig:diff_ito_pwc_pc} clearly confirms that ITO holds its precision
promise. Hence, in the following, we will assume that any propagation using ITO
is accurate to at least $\mathcal{O}\left(10^{-9}\right)$.

The results of Fig.~\ref{fig:diff_ito_pwc_pc} already indicate the strong
time-dependence of Hamiltonian~\eqref{eq:pc:ham}. Commonly, the time-dependence
of Hamiltonians can be reduced drastically by applying a rotating wave
approximation (RWA). This has exemplarily been done for the interaction
Hamiltonian~\eqref{eq:pc:ham_int}, cf. Eq.~\eqref{eq:app:H_int_RWA}. Whether
an RWA is a reasonable approximation depends, in general, on the actual system
and problem, as well as the desired accuracy. However, ITO provides an excellent
method for examining the quality of any RWA, since inaccuracies originating from
numerical propagation are drastically diminished. When repeating the
dissipation-free propagations of Fig.~\ref{fig:pop_dyn} (dashed lines) with the
RWA-Hamiltonian~\eqref{eq:app:H_int_RWA}, the average inaccuracy of the
population dynamics becomes $\bar{P}_{\text{mis}} = 1.3 \cdot 10^{-3}$. The RWA
thus turns out to be a poor approximation. Comparing the solid and dash-dotted
lines in Fig.~\ref{fig:pop_dyn} reveals moreover that, in addition to the
anharmonicity, also dissipation results in deviations from the ideal dynamics
and hampers perfect population inversion. Experimentally, Ref.
\cite{Svetitski.NatComm.5.5617} observed even more drastic discrepancies to
Pythagorean control for high intensity. We therefore analyze the dynamics under
such strong driving below.

\subsection{Dynamics under Strong Driving}
\label{subsec:pc:strong}
\begin{figure*}
  \centering
  \includegraphics{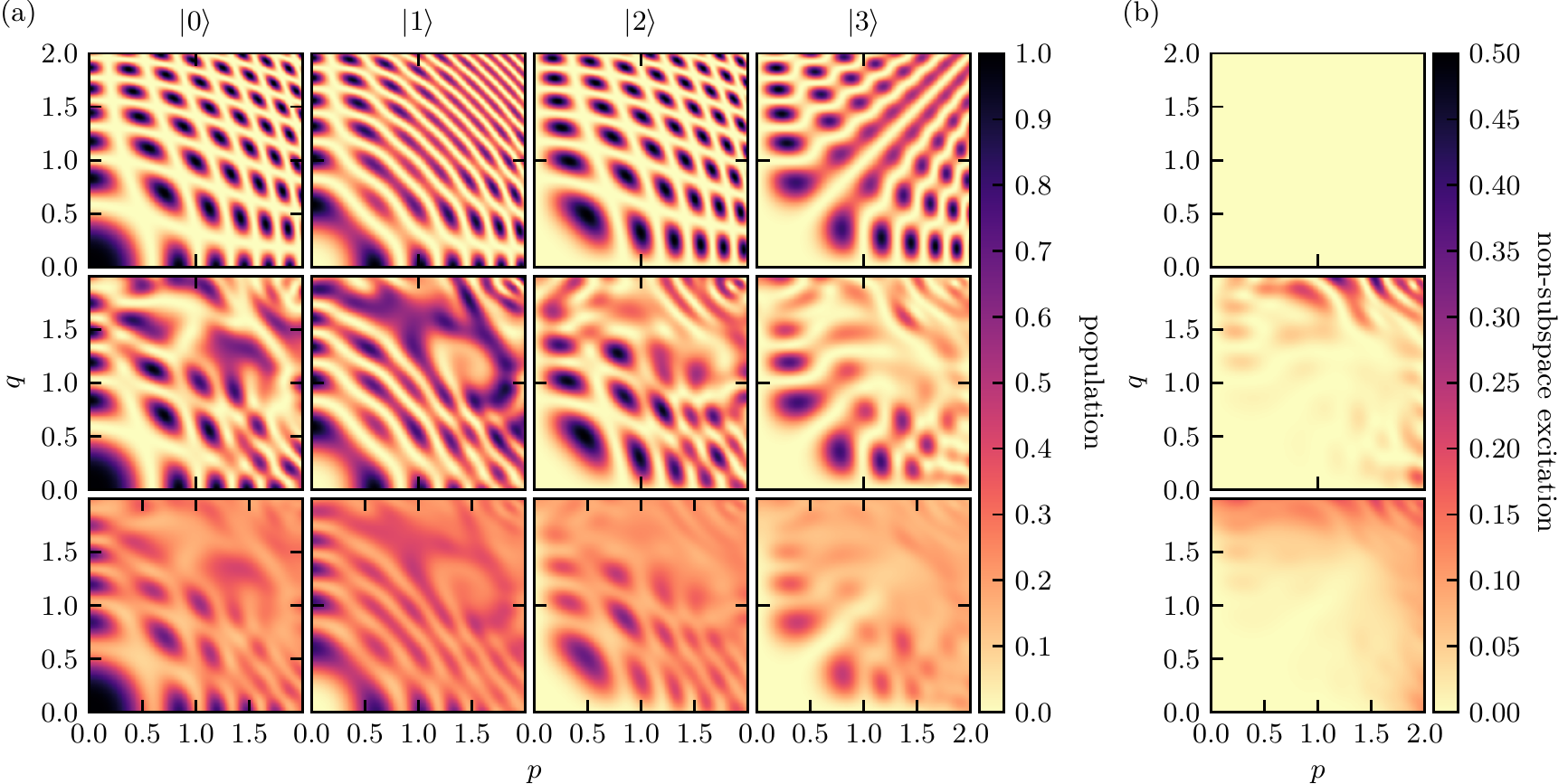}
  \caption{%
    Final populations after applying the field~\eqref{eq:pc:field} for
    $T=60$\,ns. The upper row corresponds to the ideal, non-dissipative case of
    Eq.~\eqref{eq:pc:ham_int} with only $\op{H}_{\text{inf}}$. The middle
    (lower) row shows the propagation results with
    Hamiltonian~\eqref{eq:pc:ham} neglecting (including) dissipation. All
    simulations are performed using the ITO propagator. Part
    (a) shows the final population in states $\Ket{0}, \Ket{1}, \Ket{2},
    \Ket{3}$ as indicated on top of the four columns, part
    (b) the final population outside of the subspace $\left\{\Ket{0}, \Ket{1},
    \Ket{2}, \Ket{3}\right\}$.
  }
  \label{fig:pop_map}
\end{figure*}
Figure~\ref{fig:pop_map}(a) shows the final population in the four states
$\Ket{0}$, $\Ket{1}$, $\Ket{2}$, $\Ket{3}$ after a fixed propagation time of
$T=60$\,ns, obtained with fields of various intensities. Since the field is
entirely determined by $p$ and $q$, cf. Eq.~\eqref{pc:eq:triple_scaling}, the
field intensity increases from the lower left to the upper right part within
each panel. For the ideal, non-dissipative case, depicted in the upper row of
Fig.~\ref{fig:pop_map}(a), a regular pattern can be observed. In comparison, the
middle row shows the results obtained with Hamiltonian~\eqref{eq:pc:ham}. To
emphasize the differences, dissipation has been neglected for the moment. We can
clearly identify two different regions within each map. On the one hand, for
weak field intensities, the ideal pattern is reproduced fairly well when taking
the finite anharmonicity into account. On the other hand, for strong field
intensities, the ordered structure visible in the upper row vanishes completely.
When including dissipation, all maps become blurred (lower row), since the
dissipation spreads the population across all levels. Nevertheless, the
discrepancy between weak and strong field intensities appears with and without
dissipation, i.e., the underlying effect is not related to dissipation. The
solution to the puzzle can partly be found in Fig.~\ref{fig:pop_map}(b), where
the final population outside of the subspace $L = \text{span}\left\{\Ket{0},
\Ket{1}, \Ket{2}, \Ket{3}\right\}$ is shown. In the ideal case (upper row), no
population can leave $L$, since Hamiltonian $\op{H}_{\text{inf}}$ in
Eq.~\eqref{eq:pc:ham_int} contains no coupling elements to any states $\Ket{n}$
with $n>3$. In contrast, $\op{H}_{\text{rot}}(t)$, respectively
Hamiltonian~\eqref{eq:pc:ham}, does indeed contain such couplings. As visible in
the middle and lower panels of Fig.~\ref{fig:pop_map}(b), the population leakage
out of subspace $L$ is rather small for weak fields but increases rapidly for
strong fields. Moreover, in addition to pure loss of population from subspace
$L$ at final time, the operator $\op{H}_{\text{rot}}(t)$ will also increasingly
influence the dynamics at intermediate times. In order to examine the impact of
the operator, while neglecting, at the same time, loss of population from $L$,
we truncate the qudit ladder at $N=4$. Note that this is definitively a bad
approximation, since Fig.~\ref{fig:pop_map}(b) has already shown the levels
$\Ket{n}$ with $n>3$ to contribute significantly to the dynamics. Nevertheless,
repeating the simulation of the middle row of Fig.~\ref{fig:pop_map} with $N=4$
(data not shown) yields a similar match for weak field intensities, or mismatch
for strong field intensities, respectively, compared to the ordered structure of
the ideal case. This shows that even without loss of population from the
subspace $L$, the operator $\op{H}_{\text{rot}}(t)$ compromises the ideal
population inversion of Pythagorean control.

\begin{figure*}
  \centering
  \includegraphics{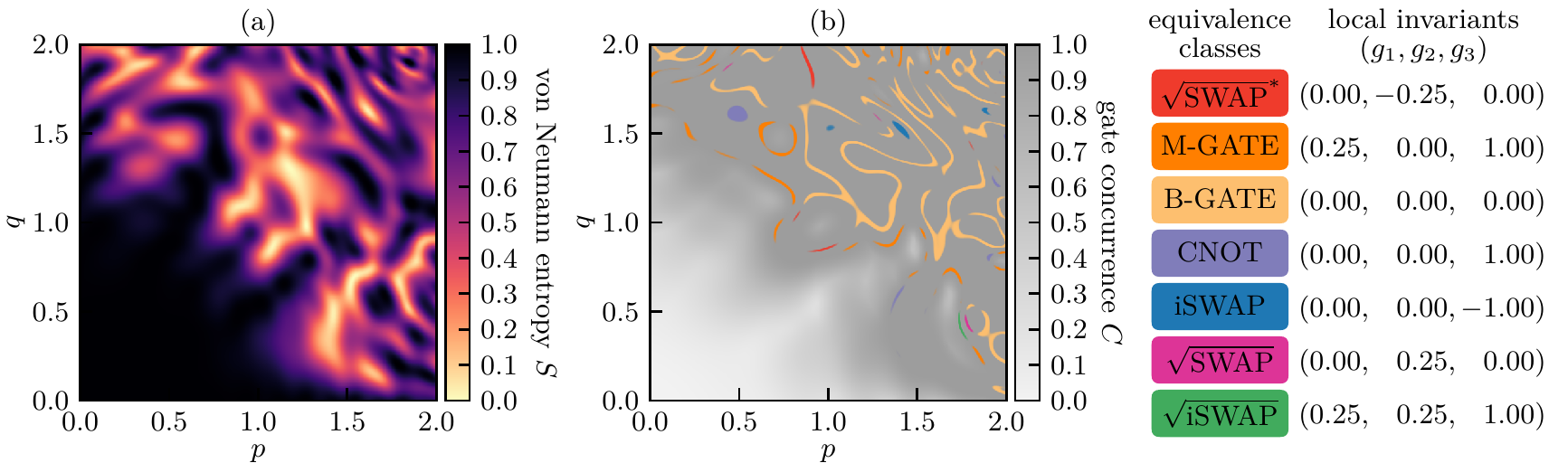}
  \caption{%
    (a) shows the von Neumann entropy $S$ for the final states of
    Fig~\ref{fig:pop_map} (middle row).
    (b) shows the gate concurrence (gray shading in the background) for the
    implemented quantum gate. The colored areas indicate where the implemented
    gates are close to the respective local equivalence classes listed on the
    right.
  }
  \label{fig:gate_map}
\end{figure*}

Interestingly, the deviation from the ideal population inversion in Pythagorean
control, caused by intense driving fields, gives rise to much richer dynamics
in terms of implementable operations, i.e., quantum gates, on the subspace
$\mathcal{L}$. As has be shown in Ref.~\cite{Suchowski.PRA.84.013414},
$\op{H}_{\text{inf}}$ allows only for quantum gates $\op{O} \in \text{SO}(4)
= \text{SU}(2) \otimes \text{SU}(2)$. One can interpret the four-level qudit as
consisting of two virtual qubits by assigning one of the four two-qubit basis
states to each of the qudit states $\Ket{0}$, $\Ket{1}$, $\Ket{2}$, $\Ket{3}$.
In this picture, it is the `non-local' operations that are missing in the ideal
case, as they are element of $\text{SU}(4) \setminus \text{SO}(4)$. According
to Ref. \cite{Svetitski.NatComm.5.5617}, we choose the four Bell states
$\{\Ket{\Phi^{\pm}}, \Ket{\Psi^{\pm}}\}$ as two-qubit basis.
Figure~\ref{fig:gate_map}(a) shows the von Neumann entropy $S$
\cite{Bennet.PRA.53.2046}, as a measure of entanglement between the two virtual
qubits, for the final states of Fig.~\ref{fig:pop_map}(a) (middle row). As can
be seen, weak fields are not able to change the amount of initial entanglement.
This perfectly agrees with the observations in Fig.~\ref{fig:pop_map}, since
weak fields yield good agreement with the prediction of Pythagorean control.
Thus, the implemented quantum gates are operations $\op{O} \in \text{SO}(4)$, at
least roughly so. For strong field intensities, the implemented quantum gates
become entangling, as the change in $S$ indicates. The gate concurrence $C$
\cite{Kraus.PRA.63.062309} can be used in order to quantify the entangling power
of these quantum gates. It ranges from $C=0$ (non-entangling) to $C=1$
(maximally entangling) and is shown in Fig.~\ref{fig:gate_map}(b) (gray
background shading) for the final states of Fig.~\ref{fig:gate_map}(a),
respectively Fig.~\ref{fig:pop_map} (middle row). As can be seen, strong field
intensities create maximally entangling gates for almost all combinations of $p$
and $q$.

We now analyze the implemented quantum gates. To this end, we perform a Cartan
decomposition, which separates each gate into its local and non-local content
\cite{Zhang.PRA.67.042313}. The non-local content unambiguously determines the
entangling power of each gate. It is given by three real numbers, the local
invariants $g_{1}, g_{2}, g_{3}$ \cite{Makhlin.QIP.1.243}. We call two gates
$\op{O}_{1}, \op{O}_{2} \in \text{SU}(4)$ locally equivalent and say they are in
the same equivalence class $[\op{O}_{1}]$, if they only differ by local
operations $\op{k}_{1}, \op{k}_{2} \in \text{SO}(4)$, i.e., $\op{O}_{1}
= \op{k}_{1} \op{O}_{2} \op{k}_{2}$. It is straightforward to calculate the
local invariants for all final gates of Fig.~\ref{fig:gate_map} and compare them
against equivalence classes of common entangling two-qubit gates, cf.\ right
column of Fig.~\ref{fig:gate_map}. For strong field intensities, we find various
regions, where the implemented gates are close to a specific equivalence class
of entangling two-qubit gates. This shows, that depending on the field
parameters $p,q$ for Pythagorean control, a large set of entangling two-qubit
gates can be realized. Note that the check for closeness to an equivalence class
is rather loose with $\left|\vec{g} - \vec{g}_{\text{ec}}\right| \leq 0.1$,
$\vec{g} = \left(g_{1}, g_{2}, g_{3}\right)$ and $\vec{g}_{\text{ec}}$
corresponding to one of the triples shown on the right side of
Fig.~\ref{fig:gate_map}. Moreover, to counter the loss of population from the
subspace $L$, which makes the actually implemented quantum gates non-unitary
within $L$, a singular value decomposition has been applied in order to get the
closest approximate unitary operation on $\mathcal{L}$. Therefore, the
characterized gates are not accurate in terms of necessary gate fidelity for
quantum computing \cite{Steane.PRA.68.042322}. Nevertheless, they are still
a hint towards the qudit's natural evolution and, in particular, they emphasize
the large amount of gates, which are accessible by varying $p,q$. Since gate
generation in Fig.~\ref{fig:gate_map} is limited to the analytical
pulse~\eqref{eq:pc:field}, it is natural to ask whether the gate fidelities can
be improved when optimizing the pulse.

\subsection{Pythagorean Control using Krotov's Method with ITO}
\label{subsec:pc:itoqoct}
We first check how Krotov's method with ITO, as described in
subsection~\ref{subsec:methods:itoqoct}, performs for the qudit under
Pythagorean control. We consider two control problems. First, in order to
compare with Sec.~\ref{sec:bench:itoqoct}, the control problem is
a state-to-state transition. Specifically, we seek to achieve population
inversion from $\Ket{0}$ to $\Ket{2}$ where it would not occur naturally. The
target functional is again given by Eq.~\eqref{eq:state2state}. Second, in order
to obtain a deeper understanding of Fig.~\ref{fig:gate_map}, the control problem
consists in implementing a predefined two-qubit gate $\op{O}$ at final time $T$.
To this end, the figure of merit $J_T$ in the optimization
functional~\eqref{eq:methods:J} is taken to be~\cite{Palao.PRA.68.062308}
\begin{align}\label{eq:pc:J_T}
  J_T [ \{ \Psi_n (T) \} ] = 1 - \frac{1}{4} \rp
    \bigg\{ \sum_{n=0}^3 \Braket{n | \op{O}^\dagger | \Psi_n (T)} \bigg\}\,.
\end{align}
The set $\{\ket{\Psi_n (t)}\}$ corresponds to the forward propagated initial
states $\{\Ket{n}\}$, given by the states from the subspace $L
= \text{span}\{\Ket{0}, \Ket{1}, \Ket{2}, \Ket{3}\}$. This control problem is
significantly more challenging than a state-to-state optimization. In terms of
control complexity, it is equivalent to optimizing simultaneously four
state-to-state optimizations, one for each state in the logical
basis~\cite{TeschPRL02,PalaoPRL02}.

\begin{figure}
  \centering
  \includegraphics{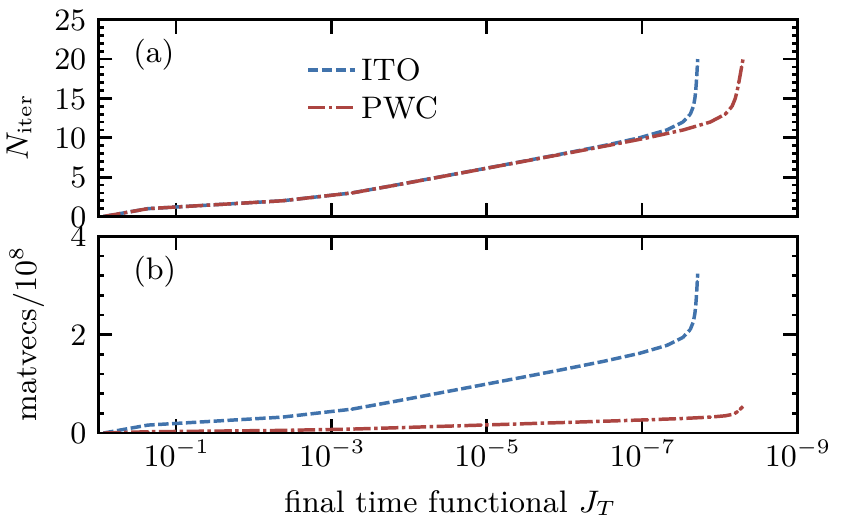}
  \caption{%
    The same comparison as in Fig.~\ref{fig:bench:krito_ho}, but for the qudit
    under Pythagorean control with the goal of realizing population inversion
    from $\Ket{0}$ to $\Ket{2}$ that does not occur naturally ($p=q=0.5$ for the
    guess pulse, $T=150\,$ns, $\delta t=0.001\,$ns, $M=6$).
  }
  \label{fig:pc:oct:state}
\end{figure}
For the first control problem, Fig.~\ref{fig:pc:oct:state} compares the
performance of Krotov's method when used with the Chebychev propagator in the
PWC approximation and the ITO propagator, respectively. While the optimization
requires the same amount of iterations to reach a certain quality of the
control, using ITO comes with a larger numerical effort, as evidenced by the
larger number of matrix vector operations in Fig.~\ref{fig:pc:oct:state}(b).
These findings are similar as those for controlling the harmonic oscillator,
reported in Fig.~\ref{fig:bench:krito_ho}. In contrast to
Fig.~\ref{fig:bench:krito_ho}, however, optimization based on the ITO propagator
cannot reach smaller values of $J_T$, i.e., more accurate controls. The
saturation of the optimization for the ITO propagator that is seen in the blue
dashed line becoming vertical, is most likely due to a rather high sensitivity
of the algorithm on the ITO parameters $\delta t$ and $M$ as well as the Krotov
update parameter $\lambda_a$, cf. Eq.~\eqref{eq:methods:eps_update}.

\begin{figure}
  \centering
  \includegraphics{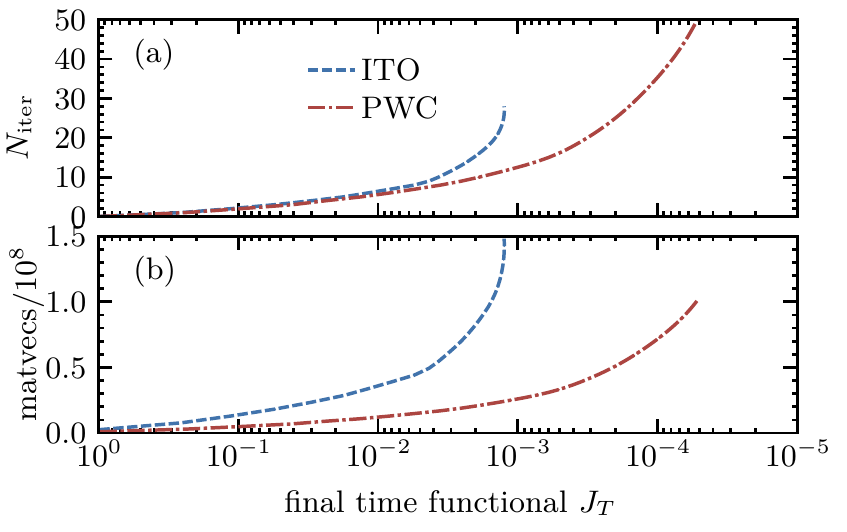}
  \caption[PWC/ITO Krotov benchmark PC]{%
    The same comparison as in Fig.~\ref{fig:bench:krito_ho}, but for the qudit
    under Pythagorean control and the CNOT gate as target operation $\op{O}$,
    cf. Eq.~\eqref{eq:pc:J_T} ($T=150\,$ns, $\delta t=0.003\,$ns, $M=8$).
  \label{fig:pc:krito_pc}}
\end{figure}
For the second control example, optimization of a CNOT gate, the difficulties of
Krotov's method when using ITO propagation become even more pronounced, see
Fig.~\ref{fig:pc:krito_pc}. The fact, that the control problem itself is more
challenging is evidenced in Fig.~\ref{fig:pc:krito_pc} by the achievable values
of $J_T$, respectively the error, being much larger than in
Fig.~\ref{fig:pc:oct:state} for both methods. But the multi-target optimization
involves yet another difficulty for the ITO propagator. The optimal choice of
its parameters $\delta t$ and $M$ must now be balanced between four different
propagations. Given the sensitivity of the method on these parameters,
convergence becomes more difficult to achieve. It turns out that smaller values
of $M$, around $M=5$ or 6, should be used for stable optimizations, in
comparison to a stable, stand-alone propagation. The optimal value of $M$
depends of course on the time step $\delta t$, which has to be chosen
accordingly (small) such that a small $M$ suffices. For maximal efficiency,
a tradeoff has to be found where the values are small enough for sufficient
stability but not so small as to be numerically too costly. But even in this
case, optimization with the ITO propagator requires significantly more
computational resources than with the PWC approximation, cf.
Fig.~\ref{fig:pc:krito_pc}.

\subsection{Controllability and Quantum Speed Limit}
\label{subsec:pc:oct}
Apart from the problem of finding optimized fields that implement a desired
dynamics, such as a specific quantum gate, OCT can be used to answer, at least
approximately, the more fundamental question of
controllability~\cite{Jurdjevic72,HuangJMP83}: When starting from a given state
$\ket{\Psi^{\text{init}}}$, which set of states $\left\{\ket{\Psi}\right\}$ is
generally accessible under the set of all possible controls? The notion of
quantum speed limits (QSLs) naturally arises in this context
\cite{Nature.406.1047, PRL.103.160502}: Provided that a specific state
$\ket{\Psi^{\text{targ}}}$ is accessible by the system's dynamics starting in
$\ket{\Psi^{\text{init}}}$, what is the minimal time for this transition? The
notion of quantum speed limit is particularly important with respect to unwanted
interaction with the environment since it tells us whether it is possible to
`beat' decoherence using optimized controls~\cite{KochJPCM16}.

\begin{figure}
  \centering
  \includegraphics{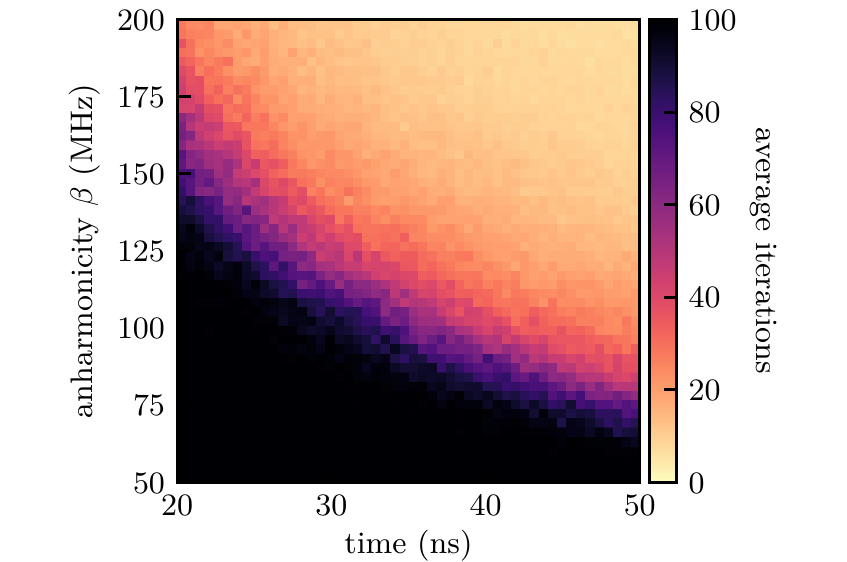}
  \caption{%
    Optimization results for random unitary operations with the optimizations
    performed for various final times $T$ and varying qudit anharmonicities
    $\beta$. The guess fields, cf. Eq.~\eqref{eq:pc:field}, with $p=q=2$, were
    shaped on input by a linear rise and fall time of $10\%$ of $T$.
    Optimizations were repeated $30$ times and stopped at $100$ iterations if
    $J_{T} < 10^{-3}$, cf. Eq.~\eqref{eq:pc:J_T}, could not be achieved until
    then.
  }
  \label{fig:qsl_map}
\end{figure}

Figure~\ref{fig:qsl_map} provides information about the qudit's controllability
in two ways. On one hand, it numerally shows full qudit controllability. On the
other hand, it allows to extract the QSL for the implementation of any two-qubit
gate. The figure shows optimization results (using Krotov's method in
combination with a PWC propagator) for a variation of the final time $T$ and the
anharmonicity $\beta$ of the qudit. Note that in order to focus on the influence
of both parameters, dissipative effects have been neglected. For each
combination of $T$ and $\beta$, $30$ random unitary gates $\op{O}$ have been
chosen \cite{Mezzadri.NotAMS.54.592} as optimization target and the average
required number of iterations in order to reach $J_{T}<10^{-3}$ is shown. The
obtained map can be clearly divided into two regions. In the lower left part of
Fig.~\ref{fig:qsl_map}, the optimization algorithm was not able reach
$J_{T}<10^{-3}$ within the allowed $100$ iterations. Most likely, it will
neither be possible for a larger number of iterations. In contrast, in the upper
right part, the optimization algorithm finds suitable fields within only a few
iterations, yielding gates with sufficiently low errors. This is numerical
evidence for full controllability as the target gates were chosen randomly. The
edge between both regions can be identified as the QSL for these operations. It
is not sharp, since a tradeoff between remaining gate error $J_{T}$ and total
time $T$ must usually be taken into account. Nevertheless, it decreases for
increasing qudit anharmonicity. For the parameters in
Tab.~\ref{tab:pc:parameters}, the QSL can be roughly identified as $T \approx
35\,\text{ns}$. In order to have faster gates, the anharmonicity $\beta$ must be
increased. Hence, increasing $\beta$ would benefit the qudit, as it reduces the
impact of dissipation by allowing for generally faster operations. Apart from
gate optimization, an increased $\beta$ would furthermore benefit the intended
population inversion of Pythagorean control, cf.\
subsection~\ref{subsec:pc:strong}, since it would diminish the strong field
deviations originating from the finite anharmonicity.

\section{Conclusions}
\label{sec:conclusions}

In order to assess the impact of time ordering in quantum optimal control, we
have combined Krotov's method with a highly efficient propagation method that
explicitly accounts for time ordering. We have tested the ensuing algorithm for
the harmonic oscillator and applied it to a superconducting circuit. For the
latter, we have also analyzed the population dynamics, starting from so-called
Pythagorean control for population inversion between non-adjacent levels in
a four-level system~\cite{Suchowski.PRA.84.013414}. For strong driving, the
dynamics of the superconducting qubit had been found experimentally to
significantly deviate from what is expected for Pythagorean
control~\cite{Svetitski.NatComm.5.5617}. We could explain this observation in
terms of higher levels in the anharmonic ladder that get populated which is
a direct consequence of the failure of the rotating wave approximation.
Furthermore, we have analyzed the time evolutions that can be generated in the
superconducting circuit by determining the type of `non-local' operations that
can be realized. This analysis suggested full controllability of the
superconducting qubit, a fact that we have confirmed by optimizing for random
unitaries and determining the quantum speed limit for each. The latter is
essentially determined by the anharmonicity.

Surprisingly, in our examples for OCT we have found the effect of time ordering
to be fairly small. Except for very small control errors, the control solutions
found within the piecewise constant approximation do not differ too much from
those obtained under explicit time ordering. For very high precision
applications as required, for example, in quantum information science below the
error correction threshold, time ordering will, however, eventually become an
effect that needs to be accounted for. At this time, iterative time
ordering~\cite{ndong2010chebychev, tal2012new, schaefer2017semi} provides the
most efficient approach to address this issue.

Iterative time ordering rewrites the explicitly time-dependent part of the
equation of motion as an inhomogeneity that needs to be determined
self-consistently. When combining this propagator with Krotov's method, it
turned out to be crucial to jointly determine both the control and the state
self-consistently to obtain a converging method. Still, the performance of the
resulting algorithm is rather sensitive with respect to its main parameters --
the expansion order of the inhomogeneity, the time step and the magnitude of the
control update. Should the algorithm be useful for practical applications, these
parameters need to be determined in a more automated way to avoid numerical
instability. This will be the subject of future work.

\begin{acknowledgments}
  We would like to thank Nadav Katz and Ronnie Kosloff for fruitful discussions.
  Financial support from the Volkswagen Foundation and from the Deutsche
  Forschungsgemeinschaft via Sonderforschungsbereich 1319 is gratefully
  acknowledged.
\end{acknowledgments}

\section*{Author Contribution Statement}
DB and LM contributed equally to this work, both in implementing code, running
simulations and writing and revising the manuscript. CPK conceived the work and
contributed to writing and revising of the manuscript.

\begin{appendix}
\section{Implementation of the ITO Propagator}
\label{app:implementation}

\begin{figure*}
  \centering
  \includegraphics{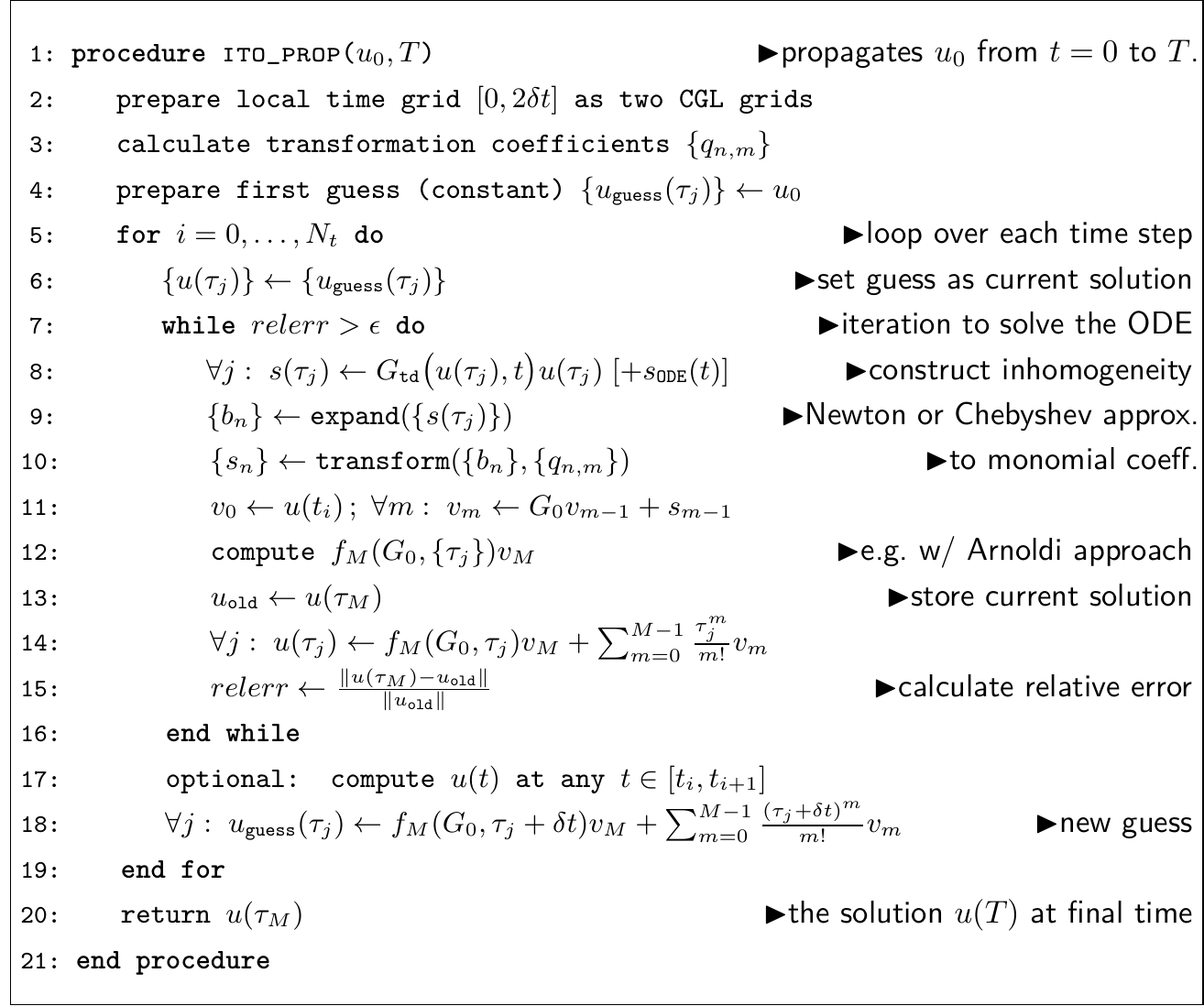}
  \caption[Pseudo code for ITO]{%
    The ITO propagator as pseudo code algorithm.
  \label{fig:methods:itoprop}}
\end{figure*}
The pseudocode for the ITO propagator is shown in
Fig.~\ref{fig:methods:itoprop}. Curved brackets indicate a loop over all
elements of the set, where the indices $i,j,m,n$ range from 0 to $M-1$. For
better readability, only the most important steps are included in
Fig.~\ref{fig:methods:itoprop}. When implementing the code for the ITO
algorithm, several aspects have to be taken into account.

(i)
Transformation of the polynomial expansion of the inhomogeneity into the
monomial basis: First, we expand Newton polynomials $R_n$ in the monomial
basis,
\begin{equation}\label{eq:app:n2t}
  R_n(t) = \sum_{m=0}^n q_{n,m} \frac{t^m}{m!}\,.
\end{equation}
Making use of the recursive definition of the Newton polynomials and by equating
coefficients, we obtain
\begin{subequations}
\begin{align}
  q_{n+1,0}   &= - t_n q_{n,0}\,, \\
  q_{n+1,m}   &= m q_{n,m-1} - t_n q_{n,m}\,, \\
  q_{n+1,n+1} &= \left( n+1 \right) q_{n,n}\,,
\end{align}
\end{subequations}
from which, with $q_{0,0} = 1$, all transformation coefficients can be computed
recursively. The inhomogeneity expanded in orthogonal Newton polynomials, with
polynomial coefficients $a_n$, can then be transformed into the monomial basis,
\begin{align*}
  s(t) \approx \sum_{m=0}^{M-1} s_m \frac{t^m}{m!}, &&
  s_m = \sum_{n=m}^{M-1} q_{n,m} a_n.
\end{align*}

(ii)
Normalization of the expansion domain: In the case of the Chebyshev polynomials
as polynomial basis, this is essential, since they can only be used on the
interval $[ -1, +1 ]$. For the Newton polynomials, a normalization also becomes
necessary when computing the coefficients by the divided difference scheme.
Because of the recursive definition, including products of the differences
between interpolating points in the denominator of the fraction, this might
become unstable depending on how small or large these differences are. The
optimal choice for stability is a domain of length $4$ which leads to a capacity
$\rho$ of size $1$~\cite{tal1991high}. This is desirable. The capacity $\rho
= \prod_{i=1}^N | z_i - z_\text{c} |^{1/N}$, where $z_\text{c}$ is the center of
the points $\{z_i\}$, is a measure for size and variance of the set $\{z_i\}$.
To obtain it, one has to take the factor $4/\delta t$ into account for the
computation of the Newton coefficients and later when using them in the
transformation to monomial coefficients, see (i).

(iii)
The product of the operator generating the dynamics and the time step, $G \delta
t$: It is recommended that this product is carried out as it is, and that
separating $G$ and $\delta t$ is avoided. As a matter of fact, the solution
contains only terms with both the operator $G$ (or its eigenvalues) and the time
step $\delta t$, neither of them individually. It might be the case that either
the eigenvalues of $G$ are large or that $\delta t$ is small, and their joint
evaluation avoids numerical instabilities.

(iv)
Calculation of the function $f_M$: An especially crucial point is the
computation of $f_M(G_0, \{\tau_j\})v_M$, where instabilities occur if it is not
computed thoroughly enough. Observing the definition of the function $f_M$, cf.
Eq.~\eqref{eq:methods:u_sol_ts}, we can see that the second term of the rhs is
just a truncated exponential sum subtracted from the full exponential function.
For small $z t$, the computation in this way might become unstable or inaccurate
due to round-off errors. Instead, it is possible to directly calculate the
expression in terms of an exponential series, starting at $M$. This decreases
the round-off error by removing the subtraction.

(v)
Error estimation for multiple sources of errors: The first -- and usually
largest -- source of error occurs in the self-consistent loop to generate the
solution, i.e., it is due to the fact that we have replaced the exact solution
$u$ by the iterated one $u^{(k)}$, cf. Eq.~\eqref{eq:methods:iter_inhom_eom}. In
order to determine how many iterations are needed, we use the common approach to
compare the new solution $u^{(k)}$ to the one from the previous iteration
$u^{(k-1)}$,
\begin{equation}
  \epsilon_{\mathrm{iter}}^{(k)} \coloneqq
    \frac{\left\| u^{(k)}(t_{n+1}) - u^{(k-1)}(t_{n+1}) \right\|}
     {\left\| u^{(k)}(t_{n+1}) \right\|}. \nonumber
\end{equation}
The second error originates from the approximation of the inhomogeneity $s(t)$
by a truncated polynomial expansion in time. This error can be estimated by
\cite{schaefer2017semi}
\begin{align}
  \epsilon_{M}
  \coloneqq
  \big\Vert
    \Delta s
  \big\Vert
  \delta t,
\end{align}
where $\Vert\Delta s\Vert$ represents the maximal interpolation error of the
approximated inhomogeneity within the current time step. For the Newton
interpolation on the Chebyshev nodes, cf. Eq.~\eqref{eq:methods:cgl_points}, it
scales as $1/(2^{M-1}M!)$. Compared to $\epsilon_\text{iter}$, it has a smaller
impact onto the total error due to the chosen splitting of the Hamiltonian,
which we chose such that the inhomogeneity is small compared to the homogeneous
part. The third and last source of error is the computation of $f_M (G_0,
\tau)v_M$, cf. Eq.~\eqref{eq:methods:u_sol_ts}. Depending on $G$, the impact of
this term might be very high in the computation. In general, if one or more of
the errors are too high, it is recommended to either increase the order of the
interpolating polynomial $M$ or decrease the size of the local interval, i.e.,
the time step $\delta t$.

\section{Derivation of the $N$-level Interaction Hamiltonian}
\label{app:derivation}
The Hamiltonian (in its generalized $N$-level form) for the superconducting
phase qudit is given by Eq.~\eqref{eq:pc:ham}. The external control field
$\pulse(t)$ is analytically given by Eq.~\eqref{eq:pc:field}. In the following,
we transform states and operators into the interaction picture. For state
$\Ket{\Psi(t)}$ in the Schr\"{o}dinger picture, the transformation reads
\begin{align} \label{eq:app:H1}
  \Ket{\Psi_{\text{int}}(t)}
  =
  \op{O}(t) \Ket{\Psi(t)},
  \qquad
  \op{O}(t)
  =
  \exp\left(\ci \op{H}_{0} t\right).
\end{align}
Plugging this into the Schr\"{o}dinger equation yields
\begin{align}
  \im \frac{\partial}{\partial t} \Ket{\Psi_{\text{int}}(t)}
  =
  \op{H}_{\text{int}}(t) \Ket{\Psi_{\text{int}}(t)}
\end{align}
with $\op{H}_{\text{int}}(t) = \op{O}(t) \op{H}_{1}(t)
\op{O}^{\dagger}(t)$. Inserting two identity operators allows to write the
interaction Hamiltonian as
\begin{align}
  \op{H}_{\text{int}}(t)
  =
  \sum_{n,m=0}^{N-1}
    \op{O}(t) \Ket{n}\Bra{n} \op{H}_{1}(t) \Ket{m}\Bra{m} \op{O}^{\dagger}(t).
\end{align}
Expanding the matrix element of $\op{H}_{1}(t)$ gives
\begin{equation} \label{eq:app:H_int_pre}
  \begin{aligned}
    \op{H}_{\text{int}}(t)
    &=
    \sum_{n,m=0}^{N-1} \sqrt{n+1} \pulse(t) \delta_{m,n+1}
      e^{- \im \omega_{n,m} t} \Ket{n}\Bra{m} + \text{H.c.}
    \\
    &=
    \sum_{n=0}^{N-2} \sqrt{n+1} \pulse(t)
      e^{- \im \omega_{n,n+1} t} \Ket{n}\Bra{n+1} + \text{H.c.}
  \end{aligned}
\end{equation}
with $\omega_{n,n+1} = \epsilon_{n+1} - \epsilon_{n}$. Expanding the analytical
field equation \eqref{eq:pc:field} in exponentials reads
\begin{equation}
  \begin{aligned}
    \pulse(t)
    &=
    \frac{1}{2} \left[%
        \frac{V_{01}}{\sqrt{1}}
          \left(e^{\im \omega_{01} t} + e^{-\im \omega_{01} t}\right)
        + \frac{V_{12}}{\sqrt{2}}
          \left(e^{\im \omega_{12} t} + e^{-\im \omega_{12} t}\right)
        \right. \\ &\qquad \left.
        + \frac{V_{23}}{\sqrt{3}}
          \left(e^{\im \omega_{23} t} + e^{-\im \omega_{23} t}\right)
      \right].
  \end{aligned}
\end{equation}
Plugging this into Eq.~\eqref{eq:app:H_int_pre}, we obtain the complete form of
the interaction Hamiltonian as
\begin{equation}
  \begin{alignedat}{1} \label{eq:app:H_int}
    \op{H}_{\text{int}}(t)
    = &
      \frac{1}{2}
      \left[
        \sum_{n=0}^{N-2} \sqrt{n+1} \Ket{n}\Bra{n+1} \times
        \right . \\ & \hspace{2mm} \left.
          \left(
              \hphantom{+} \frac{V_{01}}{\sqrt{1}} \left(
              e^{\im \left(\omega_{01} - \omega_{n,n+1}\right) t}
              + e^{- \im \left(\omega_{01} + \omega_{n,n+1}\right) t}
            \right)
          \right. \right . \\ & \hspace{4mm} \left. \left.
          + \frac{V_{12}}{\sqrt{2}} \left(
              e^{\im \left(\omega_{12} - \omega_{n,n+1}\right) t}
              + e^{- \im \left(\omega_{12} + \omega_{n,n+1}\right) t}
            \right)
          \right. \right . \\ & \hspace{4mm} \left. \left.
          + \frac{V_{23}}{\sqrt{3}} \left(
              e^{\im \left(\omega_{23} - \omega_{n,n+1}\right) t}
              + e^{- \im \left(\omega_{23} + \omega_{n,n+1}\right) t}
            \right)
          \right) \vphantom{\sum_{n=0}^{N-2}}
      \right]
      \\
      & + \text{H.c.}
  \end{alignedat}
\end{equation}
If we perform a rotating wave approximation, i.e.\ neglecting fast oscillating
terms, the interaction Hamiltonian becomes
\begin{equation}
  \begin{alignedat}{1} \label{eq:app:H_int_RWA}
    \op{H}_{\text{int}}(t)
    = &
      \frac{1}{2}
      \left[
        \sum_{n=0}^{N-2} \sqrt{n+1} \Ket{n}\Bra{n+1} \times
        \right . \\ & \hspace{2mm} \left.
          \left(
              \hphantom{+} \frac{V_{01}}{\sqrt{1}}
              e^{\im \left(\omega_{01} - \omega_{n,n+1}\right) t}
          + \frac{V_{12}}{\sqrt{2}}
              e^{\im \left(\omega_{12} - \omega_{n,n+1}\right) t}
          \right. \right . \\ & \hspace{4mm} \left. \left.
          + \frac{V_{23}}{\sqrt{3}}
              e^{\im \left(\omega_{23} - \omega_{n,n+1}\right) t}
          \right) \vphantom{\sum_{n=0}^{N-2}}
      \right] + \text{H.c.}
  \end{alignedat}
\end{equation}
Eqs.~\eqref{eq:app:H_int} and \eqref{eq:app:H_int_RWA} can be further divided
into a time-independent and time-dependent part, such that the final form
matches Eq.~\eqref{eq:pc:ham_int}.
\end{appendix}


\begin{thebibliography}{46}%
\makeatletter
\providecommand \@ifxundefined [1]{%
 \@ifx{#1\undefined}
}%
\providecommand \@ifnum [1]{%
 \ifnum #1\expandafter \@firstoftwo
 \else \expandafter \@secondoftwo
 \fi
}%
\providecommand \@ifx [1]{%
 \ifx #1\expandafter \@firstoftwo
 \else \expandafter \@secondoftwo
 \fi
}%
\providecommand \natexlab [1]{#1}%
\providecommand \enquote  [1]{``#1''}%
\providecommand \bibnamefont  [1]{#1}%
\providecommand \bibfnamefont [1]{#1}%
\providecommand \citenamefont [1]{#1}%
\providecommand \href@noop [0]{\@secondoftwo}%
\providecommand \href [0]{\begingroup \@sanitize@url \@href}%
\providecommand \@href[1]{\@@startlink{#1}\@@href}%
\providecommand \@@href[1]{\endgroup#1\@@endlink}%
\providecommand \@sanitize@url [0]{\catcode `\\12\catcode `\$12\catcode
  `\&12\catcode `\#12\catcode `\^12\catcode `\_12\catcode `\%12\relax}%
\providecommand \@@startlink[1]{}%
\providecommand \@@endlink[0]{}%
\providecommand \url  [0]{\begingroup\@sanitize@url \@url }%
\providecommand \@url [1]{\endgroup\@href {#1}{\urlprefix }}%
\providecommand \urlprefix  [0]{URL }%
\providecommand \Eprint [0]{\href }%
\providecommand \doibase [0]{http://dx.doi.org/}%
\providecommand \selectlanguage [0]{\@gobble}%
\providecommand \bibinfo  [0]{\@secondoftwo}%
\providecommand \bibfield  [0]{\@secondoftwo}%
\providecommand \translation [1]{[#1]}%
\providecommand \BibitemOpen [0]{}%
\providecommand \bibitemStop [0]{}%
\providecommand \bibitemNoStop [0]{.\EOS\space}%
\providecommand \EOS [0]{\spacefactor3000\relax}%
\providecommand \BibitemShut  [1]{\csname bibitem#1\endcsname}%
\let\auto@bib@innerbib\@empty
\bibitem [{\citenamefont {Glaser}\ \emph {et~al.}(2015)\citenamefont {Glaser},
  \citenamefont {Boscain}, \citenamefont {Calarco}, \citenamefont {Koch},
  \citenamefont {K\"ockenberger}, \citenamefont {Kosloff}, \citenamefont
  {Kuprov}, \citenamefont {Luy}, \citenamefont {Schirmer}, \citenamefont
  {Schulte-Herbr\"uggen}, \citenamefont {Sugny},\ and\ \citenamefont
  {Wilhelm}}]{GlaserEPJD15}%
  \BibitemOpen
  \bibfield  {author} {\bibinfo {author} {\bibfnamefont {S.~J.}\ \bibnamefont
  {Glaser}}, \bibinfo {author} {\bibfnamefont {U.}~\bibnamefont {Boscain}},
  \bibinfo {author} {\bibfnamefont {T.}~\bibnamefont {Calarco}}, \bibinfo
  {author} {\bibfnamefont {C.~P.}\ \bibnamefont {Koch}}, \bibinfo {author}
  {\bibfnamefont {W.}~\bibnamefont {K\"ockenberger}}, \bibinfo {author}
  {\bibfnamefont {R.}~\bibnamefont {Kosloff}}, \bibinfo {author} {\bibfnamefont
  {I.}~\bibnamefont {Kuprov}}, \bibinfo {author} {\bibfnamefont
  {B.}~\bibnamefont {Luy}}, \bibinfo {author} {\bibfnamefont {S.}~\bibnamefont
  {Schirmer}}, \bibinfo {author} {\bibfnamefont {T.}~\bibnamefont
  {Schulte-Herbr\"uggen}}, \bibinfo {author} {\bibfnamefont {D.}~\bibnamefont
  {Sugny}}, \ and\ \bibinfo {author} {\bibfnamefont {F.~K.}\ \bibnamefont
  {Wilhelm}},\ }\href {\doibase 10.1140/epjd/e2015-60464-1} {\bibfield
  {journal} {\bibinfo  {journal} {Eur. Phys. J. D}\ }\textbf {\bibinfo {volume}
  {69}},\ \bibinfo {pages} {279} (\bibinfo {year} {2015})}\BibitemShut
  {NoStop}%
\bibitem [{\citenamefont {Werschnik}\ and\ \citenamefont
  {Gross}(2007)}]{WerschnikJPB07}%
  \BibitemOpen
  \bibfield  {author} {\bibinfo {author} {\bibfnamefont {J.}~\bibnamefont
  {Werschnik}}\ and\ \bibinfo {author} {\bibfnamefont {E.~K.~U.}\ \bibnamefont
  {Gross}},\ }\href@noop {} {\bibfield  {journal} {\bibinfo  {journal} {J.
  Phys. B}\ }\textbf {\bibinfo {volume} {40}},\ \bibinfo {pages} {R175}
  (\bibinfo {year} {2007})}\BibitemShut {NoStop}%
\bibitem [{\citenamefont {Khaneja}\ \emph {et~al.}(2005)\citenamefont
  {Khaneja}, \citenamefont {Reiss}, \citenamefont {Kehlet}, \citenamefont
  {Schulte-Herbr\"uggen},\ and\ \citenamefont {Glaser}}]{KhanejaJMR05}%
  \BibitemOpen
  \bibfield  {author} {\bibinfo {author} {\bibfnamefont {N.}~\bibnamefont
  {Khaneja}}, \bibinfo {author} {\bibfnamefont {T.}~\bibnamefont {Reiss}},
  \bibinfo {author} {\bibfnamefont {C.}~\bibnamefont {Kehlet}}, \bibinfo
  {author} {\bibfnamefont {T.}~\bibnamefont {Schulte-Herbr\"uggen}}, \ and\
  \bibinfo {author} {\bibfnamefont {S.~J.}\ \bibnamefont {Glaser}},\
  }\href@noop {} {\bibfield  {journal} {\bibinfo  {journal} {J. Magn. Reson.}\
  }\textbf {\bibinfo {volume} {172}},\ \bibinfo {pages} {296 } (\bibinfo {year}
  {2005})}\BibitemShut {NoStop}%
\bibitem [{\citenamefont {Castro}\ \emph {et~al.}(2012)\citenamefont {Castro},
  \citenamefont {Werschnik},\ and\ \citenamefont {Gross}}]{CastroPRL12}%
  \BibitemOpen
  \bibfield  {author} {\bibinfo {author} {\bibfnamefont {A.}~\bibnamefont
  {Castro}}, \bibinfo {author} {\bibfnamefont {J.}~\bibnamefont {Werschnik}}, \
  and\ \bibinfo {author} {\bibfnamefont {E.~K.~U.}\ \bibnamefont {Gross}},\
  }\href {\doibase 10.1103/PhysRevLett.109.153603} {\bibfield  {journal}
  {\bibinfo  {journal} {Phys. Rev. Lett.}\ }\textbf {\bibinfo {volume} {109}},\
  \bibinfo {pages} {153603} (\bibinfo {year} {2012})}\BibitemShut {NoStop}%
\bibitem [{\citenamefont {Hellgren}\ \emph {et~al.}(2013)\citenamefont
  {Hellgren}, \citenamefont {R\"as\"anen},\ and\ \citenamefont
  {Gross}}]{HellgrenPRA13}%
  \BibitemOpen
  \bibfield  {author} {\bibinfo {author} {\bibfnamefont {M.}~\bibnamefont
  {Hellgren}}, \bibinfo {author} {\bibfnamefont {E.}~\bibnamefont
  {R\"as\"anen}}, \ and\ \bibinfo {author} {\bibfnamefont {E.~K.~U.}\
  \bibnamefont {Gross}},\ }\href {\doibase 10.1103/PhysRevA.88.013414}
  {\bibfield  {journal} {\bibinfo  {journal} {Phys. Rev. A}\ }\textbf {\bibinfo
  {volume} {88}},\ \bibinfo {pages} {013414} (\bibinfo {year}
  {2013})}\BibitemShut {NoStop}%
\bibitem [{\citenamefont {Greenman}\ \emph {et~al.}(2015)\citenamefont
  {Greenman}, \citenamefont {Koch},\ and\ \citenamefont
  {Whaley}}]{GreenmanPRA15}%
  \BibitemOpen
  \bibfield  {author} {\bibinfo {author} {\bibfnamefont {L.}~\bibnamefont
  {Greenman}}, \bibinfo {author} {\bibfnamefont {C.~P.}\ \bibnamefont {Koch}},
  \ and\ \bibinfo {author} {\bibfnamefont {K.~B.}\ \bibnamefont {Whaley}},\
  }\href {\doibase 10.1103/PhysRevA.92.013407} {\bibfield  {journal} {\bibinfo
  {journal} {Phys. Rev. A}\ }\textbf {\bibinfo {volume} {92}},\ \bibinfo
  {pages} {013407} (\bibinfo {year} {2015})}\BibitemShut {NoStop}%
\bibitem [{\citenamefont {Goetz}\ \emph {et~al.}(2016)\citenamefont {Goetz},
  \citenamefont {Karamatskou}, \citenamefont {Santra},\ and\ \citenamefont
  {Koch}}]{GoetzPRA16}%
  \BibitemOpen
  \bibfield  {author} {\bibinfo {author} {\bibfnamefont {R.~E.}\ \bibnamefont
  {Goetz}}, \bibinfo {author} {\bibfnamefont {A.}~\bibnamefont {Karamatskou}},
  \bibinfo {author} {\bibfnamefont {R.}~\bibnamefont {Santra}}, \ and\ \bibinfo
  {author} {\bibfnamefont {C.~P.}\ \bibnamefont {Koch}},\ }\href {\doibase
  10.1103/PhysRevA.93.013413} {\bibfield  {journal} {\bibinfo  {journal} {Phys.
  Rev. A}\ }\textbf {\bibinfo {volume} {93}},\ \bibinfo {pages} {013413}
  (\bibinfo {year} {2016})}\BibitemShut {NoStop}%
\bibitem [{\citenamefont {Soml\'{o}i}\ \emph {et~al.}(1993)\citenamefont
  {Soml\'{o}i}, \citenamefont {Kazakovski},\ and\ \citenamefont
  {Tannor}}]{SomloiCP93}%
  \BibitemOpen
  \bibfield  {author} {\bibinfo {author} {\bibfnamefont {J.}~\bibnamefont
  {Soml\'{o}i}}, \bibinfo {author} {\bibfnamefont {V.~A.}\ \bibnamefont
  {Kazakovski}}, \ and\ \bibinfo {author} {\bibfnamefont {D.~J.}\ \bibnamefont
  {Tannor}},\ }\href@noop {} {\bibfield  {journal} {\bibinfo  {journal} {Chem.
  Phys.}\ }\textbf {\bibinfo {volume} {172}},\ \bibinfo {pages} {85} (\bibinfo
  {year} {1993})}\BibitemShut {NoStop}%
\bibitem [{\citenamefont {Goerz}\ \emph {et~al.}(2011)\citenamefont {Goerz},
  \citenamefont {Calarco},\ and\ \citenamefont {Koch}}]{GoerzJPB11}%
  \BibitemOpen
  \bibfield  {author} {\bibinfo {author} {\bibfnamefont {M.~H.}\ \bibnamefont
  {Goerz}}, \bibinfo {author} {\bibfnamefont {T.}~\bibnamefont {Calarco}}, \
  and\ \bibinfo {author} {\bibfnamefont {C.~P.}\ \bibnamefont {Koch}},\ }\href
  {\doibase 10.1088/0953-4075/44/15/154011} {\bibfield  {journal} {\bibinfo
  {journal} {J. Phys. B}\ }\textbf {\bibinfo {volume} {44}},\ \bibinfo {pages}
  {154011} (\bibinfo {year} {2011})}\BibitemShut {NoStop}%
\bibitem [{\citenamefont {Watts}\ \emph {et~al.}(2015)\citenamefont {Watts},
  \citenamefont {Vala}, \citenamefont {M\"uller}, \citenamefont {Calarco},
  \citenamefont {Whaley}, \citenamefont {Reich}, \citenamefont {Goerz},\ and\
  \citenamefont {Koch}}]{WattsPRA15}%
  \BibitemOpen
  \bibfield  {author} {\bibinfo {author} {\bibfnamefont {P.}~\bibnamefont
  {Watts}}, \bibinfo {author} {\bibfnamefont {J.}~\bibnamefont {Vala}},
  \bibinfo {author} {\bibfnamefont {M.~M.}\ \bibnamefont {M\"uller}}, \bibinfo
  {author} {\bibfnamefont {T.}~\bibnamefont {Calarco}}, \bibinfo {author}
  {\bibfnamefont {K.~B.}\ \bibnamefont {Whaley}}, \bibinfo {author}
  {\bibfnamefont {D.~M.}\ \bibnamefont {Reich}}, \bibinfo {author}
  {\bibfnamefont {M.~H.}\ \bibnamefont {Goerz}}, \ and\ \bibinfo {author}
  {\bibfnamefont {C.~P.}\ \bibnamefont {Koch}},\ }\href {\doibase
  10.1103/PhysRevA.91.062306} {\bibfield  {journal} {\bibinfo  {journal} {Phys.
  Rev. A}\ }\textbf {\bibinfo {volume} {91}},\ \bibinfo {pages} {062306}
  (\bibinfo {year} {2015})}\BibitemShut {NoStop}%
\bibitem [{\citenamefont {Goerz}\ \emph {et~al.}(2015)\citenamefont {Goerz},
  \citenamefont {Gualdi}, \citenamefont {Reich}, \citenamefont {Koch},
  \citenamefont {Motzoi}, \citenamefont {Whaley}, \citenamefont {Vala},
  \citenamefont {M\"uller}, \citenamefont {Montangero},\ and\ \citenamefont
  {Calarco}}]{GoerzPRA15}%
  \BibitemOpen
  \bibfield  {author} {\bibinfo {author} {\bibfnamefont {M.~H.}\ \bibnamefont
  {Goerz}}, \bibinfo {author} {\bibfnamefont {G.}~\bibnamefont {Gualdi}},
  \bibinfo {author} {\bibfnamefont {D.~M.}\ \bibnamefont {Reich}}, \bibinfo
  {author} {\bibfnamefont {C.~P.}\ \bibnamefont {Koch}}, \bibinfo {author}
  {\bibfnamefont {F.}~\bibnamefont {Motzoi}}, \bibinfo {author} {\bibfnamefont
  {K.~B.}\ \bibnamefont {Whaley}}, \bibinfo {author} {\bibfnamefont
  {J.}~\bibnamefont {Vala}}, \bibinfo {author} {\bibfnamefont {M.~M.}\
  \bibnamefont {M\"uller}}, \bibinfo {author} {\bibfnamefont {S.}~\bibnamefont
  {Montangero}}, \ and\ \bibinfo {author} {\bibfnamefont {T.}~\bibnamefont
  {Calarco}},\ }\href {\doibase 10.1103/PhysRevA.91.062307} {\bibfield
  {journal} {\bibinfo  {journal} {Phys. Rev. A}\ }\textbf {\bibinfo {volume}
  {91}},\ \bibinfo {pages} {062307} (\bibinfo {year} {2015})}\BibitemShut
  {NoStop}%
\bibitem [{\citenamefont {Ndong}\ \emph {et~al.}(2010)\citenamefont {Ndong},
  \citenamefont {Tal-Ezer}, \citenamefont {Kosloff},\ and\ \citenamefont
  {Koch}}]{ndong2010chebychev}%
  \BibitemOpen
  \bibfield  {author} {\bibinfo {author} {\bibfnamefont {M.}~\bibnamefont
  {Ndong}}, \bibinfo {author} {\bibfnamefont {H.}~\bibnamefont {Tal-Ezer}},
  \bibinfo {author} {\bibfnamefont {R.}~\bibnamefont {Kosloff}}, \ and\
  \bibinfo {author} {\bibfnamefont {C.~P.}\ \bibnamefont {Koch}},\ }\href@noop
  {} {\bibfield  {journal} {\bibinfo  {journal} {J. Chem. Phys.}\ }\textbf
  {\bibinfo {volume} {132}},\ \bibinfo {pages} {064105} (\bibinfo {year}
  {2010})}\BibitemShut {NoStop}%
\bibitem [{\citenamefont {Ndong}\ \emph {et~al.}(2009)\citenamefont {Ndong},
  \citenamefont {Tal-Ezer}, \citenamefont {Kosloff},\ and\ \citenamefont
  {Koch}}]{NdongJCP09}%
  \BibitemOpen
  \bibfield  {author} {\bibinfo {author} {\bibfnamefont {M.}~\bibnamefont
  {Ndong}}, \bibinfo {author} {\bibfnamefont {H.}~\bibnamefont {Tal-Ezer}},
  \bibinfo {author} {\bibfnamefont {R.}~\bibnamefont {Kosloff}}, \ and\
  \bibinfo {author} {\bibfnamefont {C.~P.}\ \bibnamefont {Koch}},\ }\href@noop
  {} {\bibfield  {journal} {\bibinfo  {journal} {J. Chem. Phys.}\ }\textbf
  {\bibinfo {volume} {130}},\ \bibinfo {pages} {124108} (\bibinfo {year}
  {2009})}\BibitemShut {NoStop}%
\bibitem [{\citenamefont {Tal-Ezer}\ and\ \citenamefont
  {Kosloff}(1984)}]{tal1984accurate}%
  \BibitemOpen
  \bibfield  {author} {\bibinfo {author} {\bibfnamefont {H.}~\bibnamefont
  {Tal-Ezer}}\ and\ \bibinfo {author} {\bibfnamefont {R.}~\bibnamefont
  {Kosloff}},\ }\href@noop {} {\bibfield  {journal} {\bibinfo  {journal} {J.
  Chem. Phys.}\ }\textbf {\bibinfo {volume} {81}},\ \bibinfo {pages} {3967}
  (\bibinfo {year} {1984})}\BibitemShut {NoStop}%
\bibitem [{\citenamefont {Tal-Ezer}\ \emph {et~al.}(2012)\citenamefont
  {Tal-Ezer}, \citenamefont {Kosloff},\ and\ \citenamefont
  {Schaefer}}]{tal2012new}%
  \BibitemOpen
  \bibfield  {author} {\bibinfo {author} {\bibfnamefont {H.}~\bibnamefont
  {Tal-Ezer}}, \bibinfo {author} {\bibfnamefont {R.}~\bibnamefont {Kosloff}}, \
  and\ \bibinfo {author} {\bibfnamefont {I.}~\bibnamefont {Schaefer}},\
  }\href@noop {} {\bibfield  {journal} {\bibinfo  {journal} {J. Sci. Comput.}\
  }\textbf {\bibinfo {volume} {53}},\ \bibinfo {pages} {211} (\bibinfo {year}
  {2012})}\BibitemShut {NoStop}%
\bibitem [{\citenamefont {Schaefer}\ \emph {et~al.}(2017)\citenamefont
  {Schaefer}, \citenamefont {Tal-Ezer},\ and\ \citenamefont
  {Kosloff}}]{schaefer2017semi}%
  \BibitemOpen
  \bibfield  {author} {\bibinfo {author} {\bibfnamefont {I.}~\bibnamefont
  {Schaefer}}, \bibinfo {author} {\bibfnamefont {H.}~\bibnamefont {Tal-Ezer}},
  \ and\ \bibinfo {author} {\bibfnamefont {R.}~\bibnamefont {Kosloff}},\
  }\href@noop {} {\bibfield  {journal} {\bibinfo  {journal} {J. Comput. Phys.}\
  }\textbf {\bibinfo {volume} {343}},\ \bibinfo {pages} {368} (\bibinfo {year}
  {2017})}\BibitemShut {NoStop}%
\bibitem [{\citenamefont {Konnov}\ and\ \citenamefont
  {Krotov}(1999)}]{Konnov.AutomRemContr.60.1427}%
  \BibitemOpen
  \bibfield  {author} {\bibinfo {author} {\bibfnamefont {A.~I.}\ \bibnamefont
  {Konnov}}\ and\ \bibinfo {author} {\bibfnamefont {V.~F.}\ \bibnamefont
  {Krotov}},\ }\href@noop {} {\bibfield  {journal} {\bibinfo  {journal} {Autom.
  Rem. Contr.}\ }\textbf {\bibinfo {volume} {60}},\ \bibinfo {pages} {1427 }
  (\bibinfo {year} {1999})}\BibitemShut {NoStop}%
\bibitem [{\citenamefont {Sklarz}\ and\ \citenamefont
  {Tannor}(2002)}]{SklarzPRA02}%
  \BibitemOpen
  \bibfield  {author} {\bibinfo {author} {\bibfnamefont {S.~E.}\ \bibnamefont
  {Sklarz}}\ and\ \bibinfo {author} {\bibfnamefont {D.~J.}\ \bibnamefont
  {Tannor}},\ }\href {\doibase 10.1103/PhysRevA.66.053619} {\bibfield
  {journal} {\bibinfo  {journal} {Phys. Rev. A}\ }\textbf {\bibinfo {volume}
  {66}},\ \bibinfo {pages} {053619} (\bibinfo {year} {2002})}\BibitemShut
  {NoStop}%
\bibitem [{\citenamefont {Palao}\ and\ \citenamefont
  {Kosloff}(2003)}]{Palao.PRA.68.062308}%
  \BibitemOpen
  \bibfield  {author} {\bibinfo {author} {\bibfnamefont {J.~P.}\ \bibnamefont
  {Palao}}\ and\ \bibinfo {author} {\bibfnamefont {R.}~\bibnamefont
  {Kosloff}},\ }\href@noop {} {\bibfield  {journal} {\bibinfo  {journal} {Phys.
  Rev. A}\ }\textbf {\bibinfo {volume} {68}},\ \bibinfo {pages} {062308}
  (\bibinfo {year} {2003})}\BibitemShut {NoStop}%
\bibitem [{\citenamefont {Reich}\ \emph {et~al.}(2012)\citenamefont {Reich},
  \citenamefont {Ndong},\ and\ \citenamefont {Koch}}]{Reich.JCP.136.104103}%
  \BibitemOpen
  \bibfield  {author} {\bibinfo {author} {\bibfnamefont {D.~M.}\ \bibnamefont
  {Reich}}, \bibinfo {author} {\bibfnamefont {M.}~\bibnamefont {Ndong}}, \ and\
  \bibinfo {author} {\bibfnamefont {C.~P.}\ \bibnamefont {Koch}},\ }\href@noop
  {} {\bibfield  {journal} {\bibinfo  {journal} {J. Chem. Phys.}\ }\textbf
  {\bibinfo {volume} {136}},\ \bibinfo {pages} {104103} (\bibinfo {year}
  {2012})}\BibitemShut {NoStop}%
\bibitem [{\citenamefont {Kosloff}(1994)}]{RonnieReview94}%
  \BibitemOpen
  \bibfield  {author} {\bibinfo {author} {\bibfnamefont {R.}~\bibnamefont
  {Kosloff}},\ }\href@noop {} {\bibfield  {journal} {\bibinfo  {journal} {Annu.
  Rev. Phys. Chem.}\ }\textbf {\bibinfo {volume} {45}},\ \bibinfo {pages} {145}
  (\bibinfo {year} {1994})}\BibitemShut {NoStop}%
\bibitem [{\citenamefont {Breuer}\ and\ \citenamefont
  {Petruccione}(2002)}]{Breuer.book}%
  \BibitemOpen
  \bibfield  {author} {\bibinfo {author} {\bibfnamefont {H.-P.}\ \bibnamefont
  {Breuer}}\ and\ \bibinfo {author} {\bibfnamefont {F.}~\bibnamefont
  {Petruccione}},\ }\href@noop {} {\emph {\bibinfo {title} {The theory of open
  quantum systems}}},\ \bibinfo {edition} {1st}\ ed.\ (\bibinfo  {publisher}
  {Oxford University Press},\ \bibinfo {year} {2002})\BibitemShut {NoStop}%
\bibitem [{\citenamefont {Runge}\ and\ \citenamefont
  {Gross}(1984)}]{RungePRL84}%
  \BibitemOpen
  \bibfield  {author} {\bibinfo {author} {\bibfnamefont {E.}~\bibnamefont
  {Runge}}\ and\ \bibinfo {author} {\bibfnamefont {E.~K.~U.}\ \bibnamefont
  {Gross}},\ }\href@noop {} {\bibfield  {journal} {\bibinfo  {journal} {Phys.
  Rev. Lett.}\ }\textbf {\bibinfo {volume} {52}},\ \bibinfo {pages} {997}
  (\bibinfo {year} {1984})}\BibitemShut {NoStop}%
\bibitem [{\citenamefont {Marques}\ and\ \citenamefont
  {Gross}(2004)}]{MarquesARPC04}%
  \BibitemOpen
  \bibfield  {author} {\bibinfo {author} {\bibfnamefont {M.~A.~L.}\
  \bibnamefont {Marques}}\ and\ \bibinfo {author} {\bibfnamefont {E.~K.~U.}\
  \bibnamefont {Gross}},\ }\href@noop {} {\bibfield  {journal} {\bibinfo
  {journal} {Annu. Rev. Phys. Chem.}\ }\textbf {\bibinfo {volume} {55}},\
  \bibinfo {pages} {427} (\bibinfo {year} {2004})}\BibitemShut {NoStop}%
\bibitem [{\citenamefont {Marques}\ \emph {et~al.}(2012)\citenamefont
  {Marques}, \citenamefont {Maitra}, \citenamefont {Nogueira}, \citenamefont
  {Gross},\ and\ \citenamefont {Rubio}}]{TDDFT2012}%
  \BibitemOpen
  \bibinfo {editor} {\bibfnamefont {M.~A.~L.}\ \bibnamefont {Marques}},
  \bibinfo {editor} {\bibfnamefont {N.~T.}\ \bibnamefont {Maitra}}, \bibinfo
  {editor} {\bibfnamefont {F.~M.~S.}\ \bibnamefont {Nogueira}}, \bibinfo
  {editor} {\bibfnamefont {E.~K.~U.}\ \bibnamefont {Gross}}, \ and\ \bibinfo
  {editor} {\bibfnamefont {A.}~\bibnamefont {Rubio}},\ eds.,\ \href@noop {}
  {\emph {\bibinfo {title} {Fundamentals of Time-Dependent Density Functional
  Theory}}},\ \bibinfo {series} {Lecture notes in physics}, Vol.\ \bibinfo
  {volume} {837}\ (\bibinfo  {publisher} {Springer},\ \bibinfo {address}
  {Berlin, Heidelberg},\ \bibinfo {year} {2012})\BibitemShut {NoStop}%
\bibitem [{\citenamefont {Hochbruck}\ and\ \citenamefont
  {Ostermann}(2010)}]{hochbruck_ostermann_2010}%
  \BibitemOpen
  \bibfield  {author} {\bibinfo {author} {\bibfnamefont {M.}~\bibnamefont
  {Hochbruck}}\ and\ \bibinfo {author} {\bibfnamefont {A.}~\bibnamefont
  {Ostermann}},\ }\href {\doibase 10.1017/S0962492910000048} {\bibfield
  {journal} {\bibinfo  {journal} {Acta Numerica}\ }\textbf {\bibinfo {volume}
  {19}},\ \bibinfo {pages} {209–286} (\bibinfo {year} {2010})}\BibitemShut
  {NoStop}%
\bibitem [{\citenamefont {Salamin}(1995)}]{SalaminJPhysA1995}%
  \BibitemOpen
  \bibfield  {author} {\bibinfo {author} {\bibfnamefont {Y.~I.}\ \bibnamefont
  {Salamin}},\ }\href {http://stacks.iop.org/0305-4470/28/i=4/a=032} {\bibfield
   {journal} {\bibinfo  {journal} {J. Phys. A}\ }\textbf {\bibinfo {volume}
  {28}},\ \bibinfo {pages} {1129} (\bibinfo {year} {1995})}\BibitemShut
  {NoStop}%
\bibitem [{Note1()}]{Note1}%
  \BibitemOpen
  \bibinfo {note} {The computer used for all computations is an Intel Core
  i7-5930 @ 3.50\protect \tmspace +\thinmuskip {.1667em}GHz system with
  32\protect \tmspace +\thinmuskip {.1667em}GB RAM and a 64-bit Linux
  OS.}\BibitemShut {Stop}%
\bibitem [{\citenamefont {Suchowski}\ \emph {et~al.}(2011)\citenamefont
  {Suchowski}, \citenamefont {Silberberg},\ and\ \citenamefont
  {Uskov}}]{Suchowski.PRA.84.013414}%
  \BibitemOpen
  \bibfield  {author} {\bibinfo {author} {\bibfnamefont {H.}~\bibnamefont
  {Suchowski}}, \bibinfo {author} {\bibfnamefont {Y.}~\bibnamefont
  {Silberberg}}, \ and\ \bibinfo {author} {\bibfnamefont {D.~B.}\ \bibnamefont
  {Uskov}},\ }\href@noop {} {\bibfield  {journal} {\bibinfo  {journal} {Phys.
  Rev. A}\ }\textbf {\bibinfo {volume} {84}},\ \bibinfo {pages} {013414}
  (\bibinfo {year} {2011})}\BibitemShut {NoStop}%
\bibitem [{\citenamefont {Svetitski}\ \emph {et~al.}(2014)\citenamefont
  {Svetitski}, \citenamefont {Suchowski}, \citenamefont {Resh}, \citenamefont
  {Shalibo}, \citenamefont {Martinis},\ and\ \citenamefont
  {Katz}}]{Svetitski.NatComm.5.5617}%
  \BibitemOpen
  \bibfield  {author} {\bibinfo {author} {\bibfnamefont {E.}~\bibnamefont
  {Svetitski}}, \bibinfo {author} {\bibfnamefont {H.}~\bibnamefont
  {Suchowski}}, \bibinfo {author} {\bibfnamefont {R.}~\bibnamefont {Resh}},
  \bibinfo {author} {\bibfnamefont {Y.}~\bibnamefont {Shalibo}}, \bibinfo
  {author} {\bibfnamefont {J.~M.}\ \bibnamefont {Martinis}}, \ and\ \bibinfo
  {author} {\bibfnamefont {N.}~\bibnamefont {Katz}},\ }\href@noop {} {\bibfield
   {journal} {\bibinfo  {journal} {Nat. Comm.}\ }\textbf {\bibinfo {volume}
  {5}},\ \bibinfo {pages} {5617} (\bibinfo {year} {2014})}\BibitemShut
  {NoStop}%
\bibitem [{\citenamefont {Gambetta}\ \emph {et~al.}(2017)\citenamefont
  {Gambetta}, \citenamefont {Chow},\ and\ \citenamefont
  {Steffen}}]{npjQuantumInf.3.2}%
  \BibitemOpen
  \bibfield  {author} {\bibinfo {author} {\bibfnamefont {J.~M.}\ \bibnamefont
  {Gambetta}}, \bibinfo {author} {\bibfnamefont {J.~M.}\ \bibnamefont {Chow}},
  \ and\ \bibinfo {author} {\bibfnamefont {M.}~\bibnamefont {Steffen}},\ }\href
  {\doibase 10.1038/s41534-016-0004-0} {\bibfield  {journal} {\bibinfo
  {journal} {npj Quantum Inf.}\ }\textbf {\bibinfo {volume} {3}},\ \bibinfo
  {pages} {2} (\bibinfo {year} {2017})}\BibitemShut {NoStop}%
\bibitem [{\citenamefont {Reich}\ \emph {et~al.}(2015)\citenamefont {Reich},
  \citenamefont {Katz},\ and\ \citenamefont {Koch}}]{Reich.SciRep.5.12430}%
  \BibitemOpen
  \bibfield  {author} {\bibinfo {author} {\bibfnamefont {D.~M.}\ \bibnamefont
  {Reich}}, \bibinfo {author} {\bibfnamefont {N.}~\bibnamefont {Katz}}, \ and\
  \bibinfo {author} {\bibfnamefont {C.~P.}\ \bibnamefont {Koch}},\ }\href@noop
  {} {\bibfield  {journal} {\bibinfo  {journal} {Sci. Rep.}\ }\textbf {\bibinfo
  {volume} {5}},\ \bibinfo {pages} {12430} (\bibinfo {year}
  {2015})}\BibitemShut {NoStop}%
\bibitem [{\citenamefont {Bennett}\ \emph {et~al.}(1996)\citenamefont
  {Bennett}, \citenamefont {Bernstein}, \citenamefont {Popescu},\ and\
  \citenamefont {Schumacher}}]{Bennet.PRA.53.2046}%
  \BibitemOpen
  \bibfield  {author} {\bibinfo {author} {\bibfnamefont {C.~H.}\ \bibnamefont
  {Bennett}}, \bibinfo {author} {\bibfnamefont {H.~J.}\ \bibnamefont
  {Bernstein}}, \bibinfo {author} {\bibfnamefont {S.}~\bibnamefont {Popescu}},
  \ and\ \bibinfo {author} {\bibfnamefont {B.}~\bibnamefont {Schumacher}},\
  }\href@noop {} {\bibfield  {journal} {\bibinfo  {journal} {Phys. Rev. A}\
  }\textbf {\bibinfo {volume} {53}},\ \bibinfo {pages} {2046} (\bibinfo {year}
  {1996})}\BibitemShut {NoStop}%
\bibitem [{\citenamefont {Kraus}\ and\ \citenamefont
  {Cirac}(2001)}]{Kraus.PRA.63.062309}%
  \BibitemOpen
  \bibfield  {author} {\bibinfo {author} {\bibfnamefont {B.}~\bibnamefont
  {Kraus}}\ and\ \bibinfo {author} {\bibfnamefont {J.~I.}\ \bibnamefont
  {Cirac}},\ }\href@noop {} {\bibfield  {journal} {\bibinfo  {journal} {Phys.
  Rev. A}\ }\textbf {\bibinfo {volume} {63}},\ \bibinfo {pages} {062309}
  (\bibinfo {year} {2001})}\BibitemShut {NoStop}%
\bibitem [{\citenamefont {Zhang}\ \emph {et~al.}(2003)\citenamefont {Zhang},
  \citenamefont {Vala}, \citenamefont {Sastry},\ and\ \citenamefont
  {Whaley}}]{Zhang.PRA.67.042313}%
  \BibitemOpen
  \bibfield  {author} {\bibinfo {author} {\bibfnamefont {J.}~\bibnamefont
  {Zhang}}, \bibinfo {author} {\bibfnamefont {J.}~\bibnamefont {Vala}},
  \bibinfo {author} {\bibfnamefont {S.}~\bibnamefont {Sastry}}, \ and\ \bibinfo
  {author} {\bibfnamefont {K.~B.}\ \bibnamefont {Whaley}},\ }\href@noop {}
  {\bibfield  {journal} {\bibinfo  {journal} {Phys. Rev. A}\ }\textbf {\bibinfo
  {volume} {67}},\ \bibinfo {pages} {042313} (\bibinfo {year}
  {2003})}\BibitemShut {NoStop}%
\bibitem [{\citenamefont {Makhlin}(2002)}]{Makhlin.QIP.1.243}%
  \BibitemOpen
  \bibfield  {author} {\bibinfo {author} {\bibfnamefont {Y.}~\bibnamefont
  {Makhlin}},\ }\href@noop {} {\bibfield  {journal} {\bibinfo  {journal}
  {Quantum Inf. Process.}\ }\textbf {\bibinfo {volume} {1}},\ \bibinfo {pages}
  {243} (\bibinfo {year} {2002})}\BibitemShut {NoStop}%
\bibitem [{\citenamefont {Steane}(2003)}]{Steane.PRA.68.042322}%
  \BibitemOpen
  \bibfield  {author} {\bibinfo {author} {\bibfnamefont {A.~M.}\ \bibnamefont
  {Steane}},\ }\href@noop {} {\bibfield  {journal} {\bibinfo  {journal} {Phys.
  Rev. A}\ }\textbf {\bibinfo {volume} {68}},\ \bibinfo {pages} {042322}
  (\bibinfo {year} {2003})}\BibitemShut {NoStop}%
\bibitem [{\citenamefont {Tesch}\ and\ \citenamefont
  {de~Vivie-Riedle}(2002)}]{TeschPRL02}%
  \BibitemOpen
  \bibfield  {author} {\bibinfo {author} {\bibfnamefont {C.}~\bibnamefont
  {Tesch}}\ and\ \bibinfo {author} {\bibfnamefont {R.}~\bibnamefont
  {de~Vivie-Riedle}},\ }\href@noop {} {\bibfield  {journal} {\bibinfo
  {journal} {Phys. Rev. Lett.}\ }\textbf {\bibinfo {volume} {89}},\ \bibinfo
  {pages} {157901} (\bibinfo {year} {2002})}\BibitemShut {NoStop}%
\bibitem [{\citenamefont {Palao}\ and\ \citenamefont
  {Kosloff}(2002)}]{PalaoPRL02}%
  \BibitemOpen
  \bibfield  {author} {\bibinfo {author} {\bibfnamefont {J.~P.}\ \bibnamefont
  {Palao}}\ and\ \bibinfo {author} {\bibfnamefont {R.}~\bibnamefont
  {Kosloff}},\ }\href@noop {} {\bibfield  {journal} {\bibinfo  {journal} {Phys.
  Rev. Lett.}\ }\textbf {\bibinfo {volume} {89}},\ \bibinfo {pages} {188301}
  (\bibinfo {year} {2002})}\BibitemShut {NoStop}%
\bibitem [{\citenamefont {Jurdjevic}\ and\ \citenamefont
  {Sussmann}(1972)}]{Jurdjevic72}%
  \BibitemOpen
  \bibfield  {author} {\bibinfo {author} {\bibfnamefont {V.}~\bibnamefont
  {Jurdjevic}}\ and\ \bibinfo {author} {\bibfnamefont {H.~J.}\ \bibnamefont
  {Sussmann}},\ }\href {\doibase 10.1016/0022-0396(72)90035-6} {\bibfield
  {journal} {\bibinfo  {journal} {J. Diff. Eqns.}\ }\textbf {\bibinfo {volume}
  {12}},\ \bibinfo {pages} {313} (\bibinfo {year} {1972})}\BibitemShut
  {NoStop}%
\bibitem [{\citenamefont {Huang}\ \emph {et~al.}(1983)\citenamefont {Huang},
  \citenamefont {Tarn},\ and\ \citenamefont {Clark}}]{HuangJMP83}%
  \BibitemOpen
  \bibfield  {author} {\bibinfo {author} {\bibfnamefont {G.~M.}\ \bibnamefont
  {Huang}}, \bibinfo {author} {\bibfnamefont {T.~J.}\ \bibnamefont {Tarn}}, \
  and\ \bibinfo {author} {\bibfnamefont {C.~W.}\ \bibnamefont {Clark}},\
  }\href@noop {} {\bibfield  {journal} {\bibinfo  {journal} {J. Math. Phys.}\
  }\textbf {\bibinfo {volume} {24}},\ \bibinfo {pages} {2608} (\bibinfo {year}
  {1983})}\BibitemShut {NoStop}%
\bibitem [{\citenamefont {Lloyd}(2000)}]{Nature.406.1047}%
  \BibitemOpen
  \bibfield  {author} {\bibinfo {author} {\bibfnamefont {S.}~\bibnamefont
  {Lloyd}},\ }\href {\doibase 10.1038/35023282} {\bibfield  {journal} {\bibinfo
   {journal} {Nature}\ }\textbf {\bibinfo {volume} {406}},\ \bibinfo {pages}
  {1047} (\bibinfo {year} {2000})}\BibitemShut {NoStop}%
\bibitem [{\citenamefont {Levitin}\ and\ \citenamefont
  {Toffoli}(2009)}]{PRL.103.160502}%
  \BibitemOpen
  \bibfield  {author} {\bibinfo {author} {\bibfnamefont {L.~B.}\ \bibnamefont
  {Levitin}}\ and\ \bibinfo {author} {\bibfnamefont {T.}~\bibnamefont
  {Toffoli}},\ }\href {\doibase 10.1103/PhysRevLett.103.160502} {\bibfield
  {journal} {\bibinfo  {journal} {Phys. Rev. Lett.}\ }\textbf {\bibinfo
  {volume} {103}},\ \bibinfo {pages} {160502} (\bibinfo {year}
  {2009})}\BibitemShut {NoStop}%
\bibitem [{\citenamefont {Koch}(2016)}]{KochJPCM16}%
  \BibitemOpen
  \bibfield  {author} {\bibinfo {author} {\bibfnamefont {C.~P.}\ \bibnamefont
  {Koch}},\ }\href@noop {} {\bibfield  {journal} {\bibinfo  {journal} {J.
  Phys.: Condens. Matter}\ }\textbf {\bibinfo {volume} {28}},\ \bibinfo {pages}
  {213001} (\bibinfo {year} {2016})}\BibitemShut {NoStop}%
\bibitem [{\citenamefont {Mezzadri}(2007)}]{Mezzadri.NotAMS.54.592}%
  \BibitemOpen
  \bibfield  {author} {\bibinfo {author} {\bibfnamefont {F.}~\bibnamefont
  {Mezzadri}},\ }\href@noop {} {\bibfield  {journal} {\bibinfo  {journal}
  {Notices of the AMS}\ }\textbf {\bibinfo {volume} {54}},\ \bibinfo {pages}
  {592} (\bibinfo {year} {2007})}\BibitemShut {NoStop}%
\bibitem [{\citenamefont {Tal-Ezer}(1991)}]{tal1991high}%
  \BibitemOpen
  \bibfield  {author} {\bibinfo {author} {\bibfnamefont {H.}~\bibnamefont
  {Tal-Ezer}},\ }\href@noop {} {\bibfield  {journal} {\bibinfo  {journal} {SIAM
  J. Sci. Comput.}\ }\textbf {\bibinfo {volume} {12}},\ \bibinfo {pages} {648}
  (\bibinfo {year} {1991})}\BibitemShut {NoStop}%
\end{thebibliography}
%

\end{document}